\documentclass[%
 aip,
 amsmath,amssymb,
 reprint,%
]{revtex4-1}

\usepackage{graphicx}
\usepackage{dcolumn}
\usepackage{bm}

\usepackage[utf8]{inputenc}

\usepackage[T1]{fontenc}
\usepackage{mathptmx}
\usepackage{etoolbox}
\usepackage{braket}
\usepackage{accents}

\usepackage[normalem]{ulem} 

\renewcommand{\d}{\mbox{d}}
\newcommand*{\dt}[1]{%
  \accentset{\mbox{\large\bfseries .}}{#1}}

\newcommand{\rhoss}{\rho_{\rm ss}}
\newcommand{\rhoSt}{\rho (t)} 

\newcommand{\dotrhoSt}{\dt{\rho} (t)} 

\newcommand{\trB}{\mbox{tr}_\b}
\newcommand{\trS}{\mbox{tr}_\s}
\newcommand{\tr}{\mbox{tr}_{\s\b}}
\newcommand{\HS}{H_\s}
\newcommand{\Htot}{H_{\rm tot}}
\newcommand{\HB}{H_\b}
\newcommand{\Hmf}{H_\text{MF}}

\newcommand{\lamBt}{\tilde{\lambda}}

\newcommand{\tmf}{\tau_\text{MF}}
\newcommand{\stst}{\rhoss} 
\newcommand{\tauglobal}{\tau_{\s\b}}

\newcommand{\ZB}{Z_\b}
\newcommand{\Ztot}{Z_{\s\b}}
\newcommand{\Zmf}{Z_\text{MF}}
\newcommand{\D}{D}
\newcommand{\J}{J}

\newcommand{\kB}{k_\mathrm{B}}
\newcommand{\tf}{{t}} 
\newcommand{\R}{\lambda \hat{R}} 
\newcommand{\cc}{g}
\newcommand{\wk}{Weak coupling:}

\newcommand{\Dav}{{\rm D}}
\newcommand{\LD}{\tilde{\mathcal L}^{\Dav}}

\newcounter{qu} \newcommand{\QuNo}{\refstepcounter{qu}\alph{qu}}
\newcommand{\Qu}[1]{{\sf \bfseries ({#1})}}

\newcommand{\s}{{\rm S}}
\renewcommand{\b}{{\rm B}}

\usepackage{color}

\makeatletter
\def\@email#1#2{%
 \endgroup
 \patchcmd{\titleblock@produce}
  {\frontmatter@RRAPformat}
  {\frontmatter@RRAPformat{\produce@RRAP{*#1\href{mailto:#2}{#2}}}\frontmatter@RRAPformat}
  {}{}
}%
\makeatother
\begin{document}

\preprint{AIP/123-QED}

\title{Open quantum system dynamics and the mean force Gibbs state}

\author{A. S. Trushechkin}%
\affiliation{Department of Mathematical Physics, Steklov Mathematical Institute of Russian Academy of Sciences, 119991 Moscow, Russia}%
\affiliation{Department of Mathematics, National University of Science and Technology MISIS,  119049 Moscow, Russia}%

\author{M. Merkli}
\affiliation{Department of Mathematics and Statistics, Memorial University of Newfoundland, St. John's, A1C 5S7, Canada} 

\author{J. D. Cresser}
\affiliation{Department of Physics and Astronomy, University of Exeter, Exeter EX4 4QL, UK}
\affiliation{School of Physics and Astronomy, University of Glasgow, Glasgow, G12 8QQ, UK}
\affiliation{Department of Physics and Astronomy, Macquarie University, 2109 NSW, Australia}

\author{J. Anders}
\affiliation{Department of Physics and Astronomy, University of Exeter, Exeter EX4 4QL, UK}
\affiliation{Institut f\"ur Physik und Astronomie, University of Potsdam, 14476 Potsdam, Germany}

\email[]{janet@qipc.org, \quad trushechkin@mi-ras.ru, \quad merkli@mun.ca, \quad j.d.cresser@exeter.ac.uk}

\date{\today}

\begin{abstract}

The dynamical convergence of a system to the thermal distribution, or Gibbs state, is a standard assumption across all of the physical sciences. 
The Gibbs state is determined just by temperature and the system's energies alone. But at decreasing system sizes, i.e. for nanoscale and quantum systems, the interaction with their environments is not negligible. 
The question then arises: Is the system's steady state still the Gibbs state? And if not, how may the steady state depend on the interaction details?
Here we provide an overview of recent progress on answering these questions. 
We expand on the state-of-the-art along two general avenues: First we take the static point-of-view which postulates the so-called mean force Gibbs state. 
This view is commonly  adopted in the field of {\it strong coupling thermodynamics}, where modified laws of thermodynamics and non-equilibrium fluctuation relations are established on the basis of this modified state.
Second, we take the dynamical point-of-view, originating from the field of {\it open quantum systems}, which examines the time-asymptotic steady state  within two paradigms. We describe  the mathematical paradigm which proves {\it return to equilibrium}, i.e. convergence to the mean force Gibbs state, and then discuss a number of microscopic physical methods, particularly master equations. 
We conclude with a summary of established links between statics and equilibration dynamics, and provide an extensive list of open problems. 
This comprehensive overview will be of interest to researchers in the wider fields of  quantum thermodynamics, open quantum systems, mesoscopic physics, statistical physics and quantum optics, and will find applications whenever energy is exchanged on the nanoscale, from quantum chemistry and biology, to magnetism and nanoscale heat management.

\end{abstract}

\maketitle

\section{Introduction}\label{sec:intro}

Our everyday experience tells us that a macroscopic system which is brought in contact with a much larger thermal environment at temperature $T$, such as a cup of coffee in a room, itself reaches a steady state characterised by the environment's temperature. Statistical physics argues that such an equilibrium state is determined by the system energetics, given by the Hamiltonian $\HS$, as well as the temperature. Classically, this equilibrium state is known as the thermal distribution, while for quantum systems it is known as the Gibbs state 
\begin{equation} \label{eq:SGibbs}
    \tau = \frac{e^{- \HS /\kB T}}{Z},    
\end{equation} 
where $\kB$ is the Boltzmann constant and $Z$ is the partition function that normalises the density matrix $\tau$. 

Taking the Gibbs state as the equilibrium  or thermodynamically ``free'' state is a central assumption in much recent research on nanoscale and quantum thermodynamics. For example, it forms the basis of thermodynamic resource theory \cite{Aberg2013,Horodecki2013,Brandao2015,Ng2018} and is assumed in `thermal operations', which investigate the properties of CPTP maps  \footnote{Completely Positive Trace Preserving maps \cite{BrPet2002}.}  that have the Gibbs state as their fixed point \cite{Alicki2007,Nielsen2010}. 
But there is a (potentially serious) inconsistency here - the Gibbs state assumption can be problematic for exactly these `small' systems, as we will see.

\medskip

When a nanoscale or quantum system interacts with its environment, such as a molecule with the surrounding solution \cite{Jarzynski2017} or a quantum spin with the phononic modes within a material \cite{Anders2020}, the system-environment interaction energy can become comparable in size to the system's bare (or self) energy.  
This is because the surface-to-volume ratio of smaller systems is much higher than that of macroscopic systems, e.g. scaling as $R^2/R^3 = 1/R$ for a spherical system with radius $R$. For short range interactions, the system's surface size determines the strength of interaction with its environment, while the system's self-energy usually scales with volume. This implies that, while negligible for macroscopic systems, the interaction energy is relevant for systems  of decreasing size \cite{Jarzynski2017,Strasberg2017,Miller2018,Anders2020}.

Conventional Gibbs state statistical physics, which makes the tacit assumption that this interaction is vanishingly weak, see Fig.~\ref{fig:weakvsstrong},  then no longer applies. This motivates the core questions addressed in this overview article. Let  $\rho(t)$ be the system density matrix at time $t$ and denote the system steady state as
\begin{equation}
    \label{convtoss}
\stst := \lim_{t \to \infty}\rho(t).
\end{equation}
We ask:
\begin{itemize}
\item[(Q)]  Is the system steady state $\stst$ the Gibbs state $\tau$? \\
If not, how does $\stst$ depend on the interaction details?
\end{itemize}

\begin{figure}[t]
\flushleft 
\includegraphics[trim=0cm 0cm 0cm 0cm,clip, width=0.48\textwidth]{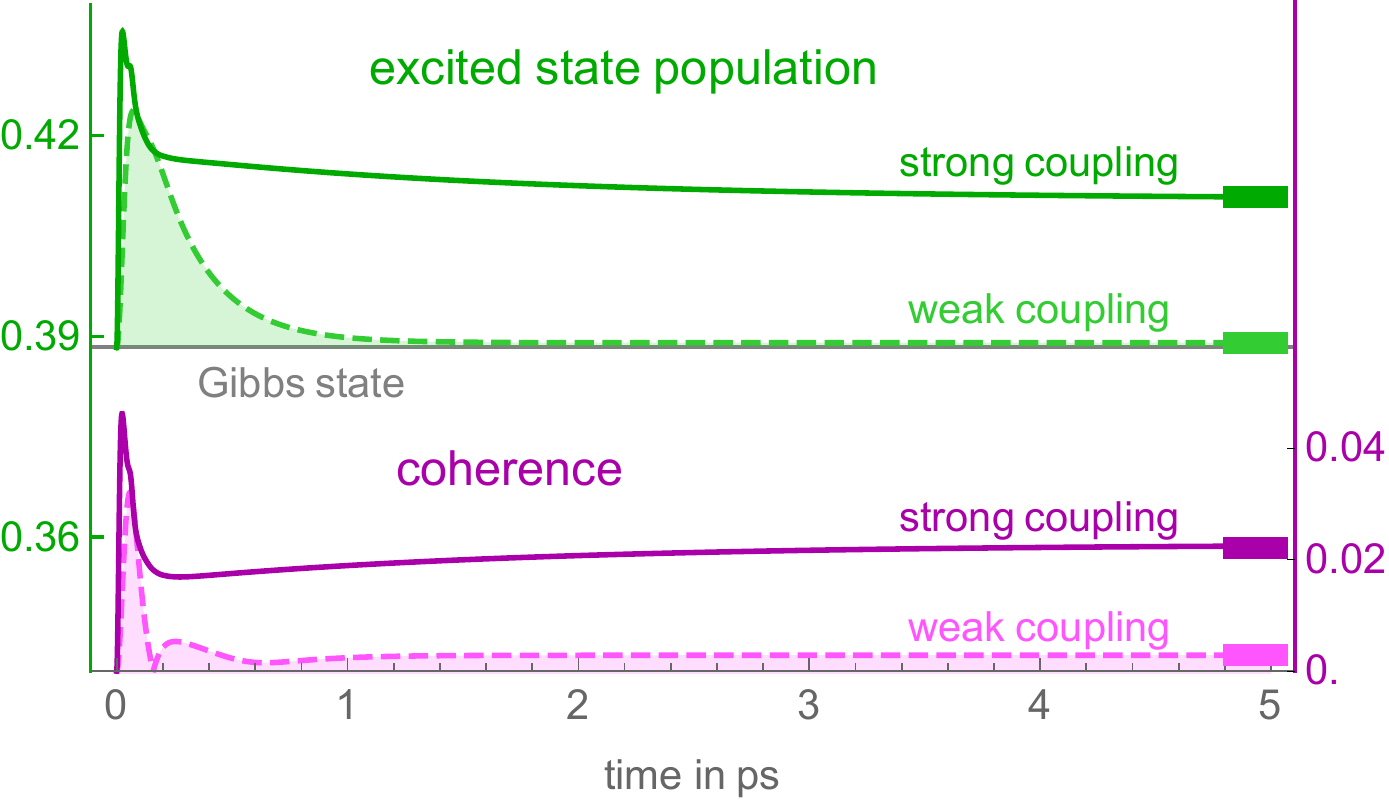}
\caption{ 
Illustration of relaxation dynamics to steady state (boxes) for a qubit that interacts weakly (interaction strength = $0.4 \times 10^{-21}~\rm{J}$) or strongly ($4 \times 10^{-21}~\rm{J}$) with a bath at temperature $T=317~\rm{K}$. 
The  energy gap between  the qubit ground state $\ket{g}$ and the excited state $\ket{e}$ is $2 \times 10^{-21}~\rm{J}$. 
For each coupling strength, the excited state population $\braket{e|\rho (t)|e}$ (green) and the absolute coherence  $\left|\braket{e|\rho(t)|g}\right|$  (magenta) are plotted as a function of time $t$. 
The bath relaxation time is $0.1~\rm{ps}$.
Deviations from the Gibbs state (excited state population is $0.388$ and coherence is 0) are shaded for the weak coupling case, and clearly much larger at larger coupling. 
Further details on the plot parameters are given in \cite{figurefootnote}.}
\label{fig:equilibration}
\end{figure}

An illustration of the dynamical approach to a stationary state, as in  \eqref{convtoss}, is provided in  Fig.~\ref{fig:equilibration} for a two-level system.

As we will see,  relaxation without recurrences is an irreversible effect which, mathematically speaking, can only happen if the dynamics has no `eternal oscillations'. Such convergence to a steady state is attributed to the environment. However, not all environments cause such irreversible effects. For example,  an environment consisting of a single qubit cannot make another qubit relax to a steady state. We will call an environment which {\em can} induce the convergence of the system (\s) to a steady state a ``{\em bath}'' (\b). Baths must have certain properties (large size, infinitely many degrees of freedom, a continuum of energies...) which will be detailed in Sec.~\ref{sec:dynamicsA}.

\subsection{Dynamic and static points of view} \label{subsec:dyn+stat}

There are two points of view to answering the questions (Q):  
In the {\it dynamic} point-of-view, one considers the dynamics of the  system which continuously interacts with an environment, beginning from an initial state $\rho_{\s\b} (0)$.  One then asks whether the reduced state of the system alone, $\rhoSt := \trB[\rho_{\s\b}(t)]$, stops changing (significantly) at late times $t$. If this is so, then one may define (one or more) system steady states $\stst$.  It is often assumed that the initial $\s\b$ state is uncorrelated, $\rho_{\s\b}(0)=\rho (0)\otimes \tau_\b$, where
$\tau_\b := {e^{-\HB/\kB T} / \trB[e^{-\HB/\kB T}]}$  is the bath Gibbs state (bath Hamiltonian $\HB$, temperature $T$). Furthermore it is common to make the so-called {\em Born approximation},  $\rho_{\s\b}(t)\approx \rhoSt \otimes\tau_\b$ for all times $t\ge0$, implying that any effect of the system on the reduced bath state can be neglected, as well as any correlations that may built up between $\s$ and $\b$.

In the {\it static} point-of-view, one postulates that at the end of any equilibration process, the combined system+bath complex is in the global Gibbs state $\tauglobal \propto e^{- \Htot /\kB T}$, where $\Htot$ is the total (interacting) Hamiltonian and the temperature is the same as the bath's temperature at the beginning of the equilibration process. The system equilibrium state is then ``simply'' the reduced state of the global state, $\tmf = \trB[\tauglobal]$, called the {\em mean force Gibbs state}.

Each avenue has its merits and limitations. 
For instance, many analytical results from the dynamics point-of-view are  based on the analysis of master equations (MEs), which are approximations of the true system dynamics. MEs are a powerful and widely used tool for the assessment of dissipation of quantum systems. 
The applicability of most MEs requires weak  -- whilst not negligible (see Fig.~\ref{fig:weakvsstrong}) -- system-bath interaction strengths and additional further  approximations. 
Different approximations can lead to different steady states $\stst$.  
Furthermore, quite often, the approximations are {\it chosen} such that the ME's steady state becomes the standard Gibbs state $\tau$, turning the question regarding the steady state somewhat on its head.

On the other hand, the static point-of-view has given rise to the subfield of {\it strong coupling thermodynamics}, see Fig.~\ref{fig:weakvsstrong},  concerned with building a thermodynamic framework that correctly includes the bath's fingerprint 
\cite{Jarzynski2004a,Campisi2009a,Campisi2009b,Gelin2009,Hilt2011,Hilt2011a,Williams2011, Seifert2016,Strasberg2016,Philbin2016,Bruch2016,Jarzynski2017,Aurell2017,Strasberg2017b, Newman2017, Miller2017,Correa2017,Aurell2018,Miller2018,Schaller2018,Strasberg2018b,Miller2018a,Dou2018,Strasberg2019,Hovhannisyan2018a,Perarnau-Llobet2018,Strasberg2019a,Huang2020,Rivas2020,Strasberg2020a}.
Based on the definition of an effective system Hamiltonian in equilibrium, called the mean force Hamiltonian, a general theory has been established, which includes thermodynamic laws and stochastic fluctuation relations for out-of-equilibrium processes. While there is some debate about the non-uniqueness of the mean force Hamiltonian, the mean force Gibbs state $\tmf$ is generally accepted as the (unique) formal equilibrium state.  An explicit expression of this state in terms of system operators alone is, however, only known in a handful of cases. This leaves much of the difficulty of including the environment in the reduced system state unresolved.
Moreover, answering whether and/or when the dynamical steady state(s) $\stst$ and the reduced global equilibrium state $\tmf$ are identical has been addressed only  relatively recently for a handful of settings. 

In this article, we assemble results -- originally reported in many individual papers -- into a focused, comparative overview
describing both the static and the dynamic points-of-view. 
We begin with detailing expressions for the static mean force Gibbs state  (MFG state) in section \ref{sec:statics}, followed by discussing mathematical results on the dynamical Return to Equilibrium (RtE) in  section \ref{sec:dynamicsA}. Section~\ref{sec:dynamicsB} gives a brief summary of key results on the dynamical steady states of microscopic master equations and other dynamical methods, and how these compare to the MFG state. We conclude in Sec.~\ref{sec:conclusions} with a summary of the outlined  state-of-the-art on the link between dynamics and statics, and end with a (rather long) list of open questions.

\smallskip

Before embarking on the above topics, we  will first set out the general setting of open quantum systems and clarify the naming convention we will use for various coupling regimes.

\subsection{General setting and Coupling strength regimes} 

The starting point for describing an open systems is to view it as a subsystem of a bigger, {\it closed} bipartite system $\s\b$, consisting of the system and the remaining part, called the bath. 
Deciding which part of an interacting complex is $\s$ and which is $\b$ may seem somewhat arbitrary.
Intuitively, the system is understood to consist of the degrees of freedom that one can manipulate and/or measure, such as the position and momentum of a pendulum, while the bath consists of degrees of freedom (DoFs) that are uncontrolled, such as the air molecules that dampen the pendulum's motion. 
The two components $\s$ and $\b$ are not equal partners: the bath influences the system's thermodynamic and dynamical properties significantly, while $\s$  cannot move $\b$ too far from its initial state. This is modeled by taking bath Hamiltonians $H_\b$ which have a continuum of energies, while system Hamiltonians $H_\s$ have only discrete energies. In models where the bath has a spatial structure (say, the bath consists of a gas of quantum particles allowed to move in a region $R\subset {\mathbb R}^3$ of position space), continuous bath energies arise in the limit of infinite volume ($R\rightarrow {\mathbb R}^3$), see also Section \ref{sec:continmodes}.

The total Hamiltonian of a general $\s\b$  complex has the form
\begin{equation}\label{eq:Htot}
	\Htot =\HS + \HB + \lambda \, V_{\s\b}, 
\end{equation}
where $\lambda\in \mathbb R$ is a dimensionless coupling constant, and $V_{\s\b}$ is the system-bath interaction operator. The latter is generally of the form
\begin{equation}\label{vint}
    V_{\s\b} =  \sum_j X^{(j)}\otimes B^{(j)},
\end{equation}
where the $X^{(j)}$ and $B^{(j)}$ are operators acting on the system and bath Hilbert spaces, respectively. 

Note, that throughout the text, we will subindex states for system+bath with ${}_{\s\b}$, {\em e.g.}, the global Gibbs state is $\tau_{SB}$, and we will subindex states for the bath alone with ${}_{\b}$, but for the system states we will drop the index ${}_\s$, {\em e.g.}, the system  Gibbs state is denoted $\tau$. We will however keep the index $S$ for the system Hamiltonian $\HS$. Furthermore, unless otherwise stated, we set $\hbar=1$.

\medskip

One distinguishes several regimes related to the magnitude of the coupling strength $\lambda$. Usually these are set by a comparison of typical system ($\HS$) energy differences and energies associated with the interaction $V_{\s\b}$.  
We note that the current naming convention used in the theory of open quantum systems literature and the strong coupling thermodynamics literature are not uniform. 

\begin{figure}[t]
    \includegraphics[trim=1.3cm 0.2cm 2.5cm 1.0cm,clip, width=0.48\textwidth]{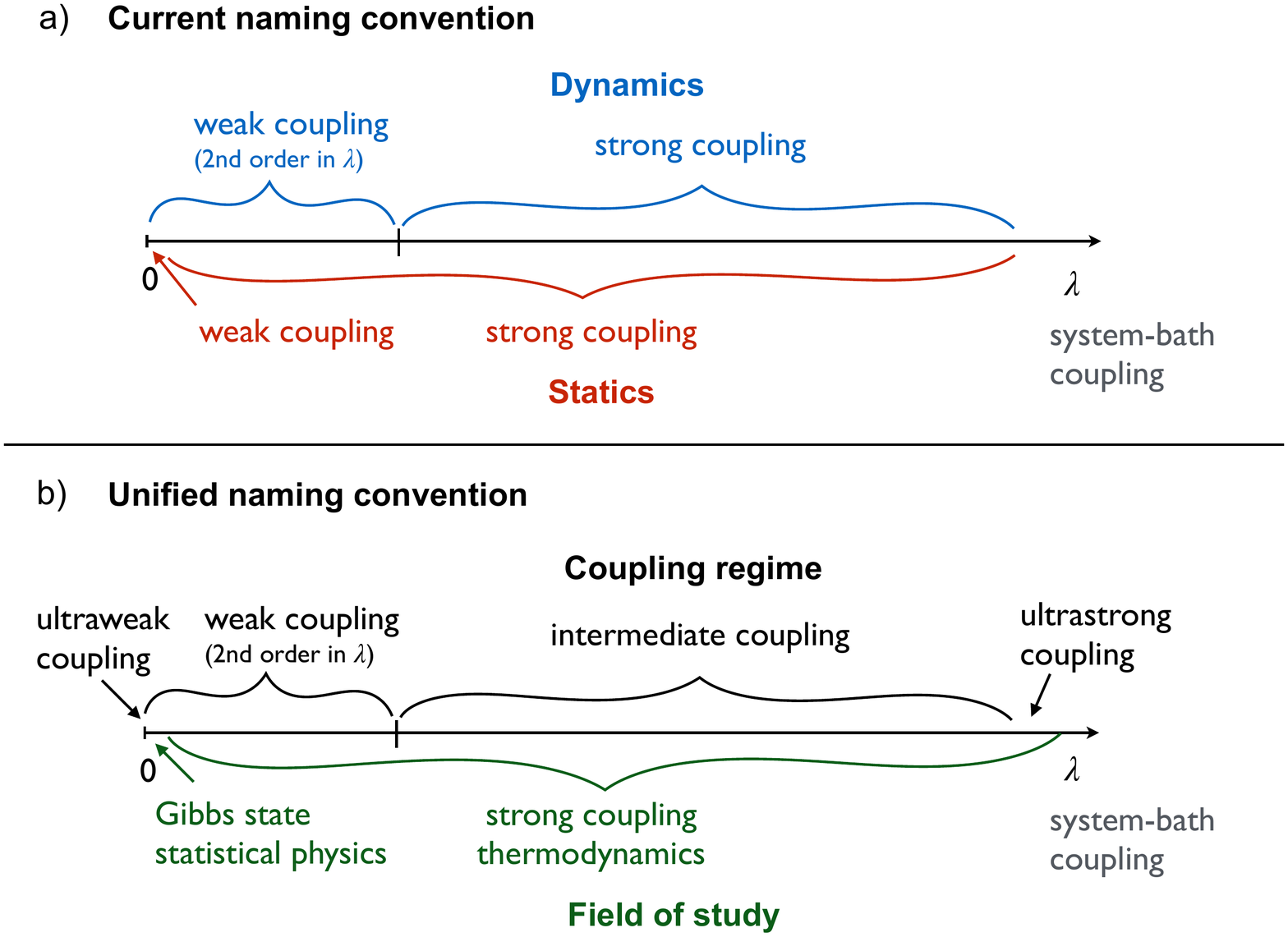}
	\caption{ {\sf a)} Coupling regime conventions commonly used in the current theory of open quantum systems literature (Dynamics) and the strong coupling thermodynamics literature (Statics). 
	Note that the dynamical ``weak coupling regime'' is contained within the field of ``strong coupling thermodynamics''. \quad 
	{\sf b)} Proposed unified naming convention for various coupling regimes and fields of study.}
	\label{fig:weakvsstrong}
\end{figure}

Here we propose a unified naming of the regimes, see Figure \ref{fig:weakvsstrong} for a visual illustration.  Our definition of the regimes is as follows: 
The {\em weak coupling regime} describes the regime where $\lambda$ is small, and perturbation theory in $\lambda$, usually to second order in $\lambda$, is justified. Weak coupling regime examples include many master equation derivations, see Sec.~\ref{subsec:weakdyn}. 
The {\em ultraweak coupling regime} is taken when all terms of order  $\lambda$ and higher are neglected. 
In the other extreme, the {\em ultrastrong coupling regime} is 
achieved when $\lambda$ is very large,  perturbation theory in $\lambda^{-1}$ may be performed, and all orders $\lambda^{-1}$ and higher can be neglected.  
In the {\em intermediate coupling regime}, $\lambda$ cannot be considered either  large or small. In this challenging regime, either non-perturbative methods or other kinds of approximations are required.

\section{Statics} \label{sec:statics}

Here we discuss the static point-of-view which arises from equilibrium statistical mechanics. 

The classical Boltzmann distribution as well as the quantum Gibbs state $\tau$ are justified by a number of arguments \cite{Balian2007}. Gibbs' original derivation \cite{Gibbs} (see also \cite{LandauLifshitzStat,Kubo1957,Khinchin,Balian2007}) considers the sharing of energies between two systems in equilibrium, and makes use of the equal probability postulate for microstates. 
Many modern approaches to deriving the canonical equilibrium state in the classical and quantum regime  follow the maximum entropy principle \cite{Jaynes1957,Jaynes1957a}. 
This approach is justified by the second law of thermodynamics which introduces the concept of irreversible entropy production, and with it a direction towards higher entropy states. 
{\it Gibbs state statistical physics}, see Fig.~\ref{fig:weakvsstrong}, emerges when the equilibrium state of a system with fixed energy $\langle U \rangle$ at temperature $T$ is taken to be the state which maximizes entropy, under the fixed energy constraint. 
To evaluate this maximum, one needs to know the energy operator, i.e. a system Hamiltonian $\HS$, and chose an expression for the entropy, usually taken to be the Shannon or von Neumann entropy \cite{Neumann1927}. 
An implicit assumption is that neither the Hamiltonian nor the entropy functional depend on properties of the equilibrium state, such as the temperature.
Maximization introduces a Lagrange multiplier $\beta$ which, upon equating the average statistical energy with $\langle U \rangle$, becomes $\beta=1/\kB T$. The resulting state of maximum entropy  takes the form of the {\it Gibbs state} $\tau$.

Thermodynamically, the system is postulated to reach this equilibrium state when it has been in weak thermal contact with a bath at temperature $T$ for a long enough time. Within Gibbs state statistical physics,  no further explicit mention is made of any bath.  The only impact the bath is assumed to have on the system is that it determines the system's temperature $T$, and that the energy of the system is subject to (statistical) fluctuations around the fixed mean value $\langle U \rangle$.

\subsection{Mean Force Gibbs state $\tmf$}  \label{subsec:MFGstate}

We now consider the system and bath complex, $\s\b$, to be in the global  Gibbs  state $\tauglobal$ associated to the total Hamiltonian $\Htot$, \eqref{eq:Htot}.
The emergence of this state can be justified -- for now -- by considering that the $\s\b$ compound has been in thermal contact for a long time with a {\em super-bath} \cite{Campisi2009b,Kosloff2019} $R$ at inverse temperature $\beta=1/\kB T$.  Gibbs state statistical physics for the compound $\s\b$ then tells us that the equilibrium state for $\s\b$ is the Gibbs state
\begin{equation}\label{eq:tausb}
	\tauglobal :=\frac{e^{-\beta \Htot}}{\Ztot},
\end{equation}
where $\Ztot=\tr[e^{-\beta \Htot}]$ is the global partition function. 
The {\em mean force Gibbs state} (MFG state) is defined as the system state obtained by taking the partial trace over the bath degrees of freedom,
\begin{equation}\label{eq:MFGibbs}
	\tmf := \trB[\tauglobal].
\end{equation}
Generally, $\tmf$ differs -- sometimes substantially -- from the Gibbs state $\tau\propto e^{-\beta H_\s}$, as we will see below. The naming arises from casting $\tmf$ in Gibbsian (exponential) form,
\begin{equation}\label{eq:MFGibbsH}
	\tmf = : \frac{e^{-\beta \Hmf}}{\Zmf} ,
\end{equation}
for an effective Hamiltonian $\Hmf$, called the Hamiltonian of mean force (HMF) or the potential of mean force. Unlike the bare system Hamiltonian $\HS$ used throughout standard statistical physics, the HMF is temperature dependent. It will also depend on the coupling strength $\lambda$ and some of the details of the interaction $V_{\s\b}$ in \eqref{eq:Htot}. Note that $\Hmf$ is not uniquely defined 
but nevertheless, the MFG state $\tmf$, \eqref{eq:MFGibbs}, is uniquely defined. 
The MFG state, as well as the HMF, have found widespread use in chemistry  \cite{Kirkwood1935,Roux1995,Roux1999,Maksimiak2003,Allen2006,Lahey2020,Wang2020} 
since the 1930s. 

Before proceeding with the discussion of the  state $\tmf$, let us first comment on the HMF. 
Unlike the bare system Hamiltonian $\HS$ used throughout Gibbs state statistical physics,  the HMF is temperature dependent. (As a consequence, identities of statistical physics will not necessarily hold and require corrections.)
It will also depend on the coupling strength $\lambda$ and some of the details of the interaction $V_{\s\b}$ in \eqref{eq:Htot}. 
Note that the $\Hmf$ is not uniquely defined since, e.g., a constant may be added to it without changing $\tmf$  in \eqref{eq:MFGibbsH}. 
This is because such constant would also change the partition function $\Zmf = \trS[e^{-\beta \Hmf}]$. (The constant also cancels when calculating energetic differences). 
A common choice is to set $\Zmf = \Ztot/\ZB$ with $\ZB$ the bare bath partition function, and to include certain strong coupling corrections into energetic and entropic potentials  \cite{Seifert2016,Jarzynski2017, Miller2017,Strasberg2017,Miller2018,Strasberg2020a}. This leads to an extensive (additive) behaviour of the effective system and bare bath potentials for classical and quantum systems, which mirrors that of standard thermodynamics.
Other thermodynamically consistent choices are being discussed \cite{Jarzynski2017,Strasberg2020a} and approaches to determining the physically meaningful HMF are being explored \cite{Strasberg2020a,TalknerComment2020,StrasbergReply2020}.

Meanwhile, based on the above definitions, much progress has been made in constructing a comprehensive framework of ``strong coupling thermodynamics" \cite{Miller2018} that includes corrections arising from the system's interaction with the environment. 
Strong coupling thermodynamic potentials have  been identified \cite{Seifert2016,Philbin2016,Jarzynski2017,Aurell2017,Aurell2018}, detailed entropy fluctuation relations have been shown to hold \cite{Strasberg2017, Miller2017}, and the validity of the Jarzynski equality \cite{Jarzynski2004,Campisi2009a,Campisi2009b} and the Clausius inequality \cite{Gelin2009,Hilt2011a,Hilt2011} have been proven. 
The strong coupling impact on a Maxwell demons' operation has been elucidated \cite{Schaller2018,Strasberg2018b}, an extension of Bohr's energy-temperature uncertainty relation to the strong coupling limit has been proven \cite{Miller2018a}, and quantum measurements have been included  in a stochastic description of strongly coupled quantum systems \cite{Strasberg2019}. 
In quantum thermometry, strong coupling has been found to  improve measurement precision \cite{Correa2017,Hovhannisyan2018a}, while it can be detrimental for the efficiency of quantum engines \cite{Perarnau-Llobet2018}.

We now return to the MFG state $\tmf$ which is uniquely defined by the formal identity \eqref{eq:MFGibbs}. 
But unfortunately, giving explicit expressions for $\tmf$ in terms of system operators alone is very often  intractable -- because it requires carrying out the trace over the (large number of) bath DoFs. Exact results have been obtained for the quantum harmonic oscillator interacting with a bath of oscillators, see \ref{subsec:qho}. For more general systems $\s$, still interacting with a bosonic bath, perturbative results have been established in the weak coupling limit,  see \ref{subsec:staticweak},  as well as the ultrastrong coupling limit, see \ref{subsec:staticultrastrong}.

\subsection{Open systems with discrete bosonic environment} 
\label{subsec:discbosonic}

The paradigmatic open quantum system model is a system with Hamiltonian $\HS$ coupled to a field of quantum harmonic oscillators according to the Hamiltonian \eqref{eq:Htot} with \cite{BrPet2002, Rivas2011}
\begin{equation}\label{m1}
    \HB = \sum_{k}\omega_k \, a^\dagger_k a_k, \quad V_{\s\b}=  X \otimes \sum_{k} \cc_k \, a^\dag_k+{\rm h.c.},
\end{equation}
where $\omega_k$ are the frequencies of the oscillator (modes) $k$, the creation $a^\dag_k$ and annihilation operators $a_k$ obey the commutation relations $[a_k , a^\dag_\ell] =\delta_{k,\ell}$, and $X$ is an arbitrary system operator. The $\cc_k$ are complex numbers that weigh the strength of the interaction between $\s$ and the oscillator mode at frequency $\omega_k$. Even though $H_\b$, \eqref{m1}, has infinitely many energy levels, those levels do not fill a continuum of values. So technically, according to our bath definition, $H_\b$ is {\em not} the Hamiltonian of a `bath'. Nevertheless, the discrete mode model \eqref{m1} often serves as a starting point, see also Section \ref{subsec:contlimit}.
An additional, so-called counter term $\lambda^2 \sum_{k} |\cc_k|^2 \, X^2/\omega_k$ is often included in the Hamiltonian, which physically arises whenever coupling is introduced via the difference of coordinates, for example  $(x_k - X)^2$, instead of a product, e.g. $x_k X$.
When included, the total Hamiltonian is  
\begin{multline}\label{eq:Hwcounter}
    \HS + \HB + \lambda \, V_{\s\b}+\lambda^2 \sum_{k} \frac{|\cc_k|^2}{\omega_k} \, X^2 \\
    = \HS + \sum_{k} \omega_k \left( a^\dag_k   +  \frac{\lambda \cc_k}{\omega_k} \, X  \right) \left(a_k +  \frac{\lambda \cc_k^{*}}{\omega_k} \, X  \right),
\end{multline}
where the bath oscillators are now {\it displaced} by the system operators. 
Note that we here neglect the zero-point energy  of the bath  oscillators, i.e. $\sum_k\omega_k/2$, which cancels in the reduced system state. This energy diverges in the continuum mode limit, and dropping it amounts to a renormalization of the bath energy.

When the system is a two level system, then Eq.~\eqref{m1} is the Hamiltonian of the ubiquitous {\it spin-Boson model}, used to describe a wide range of physical systems \cite{Thoss2001,BrPet2002,Anders2007,Boudjada2014,Yang2014,Nazir2016,Purkayastha2020}, ranging from qubits in quantum computers\cite{Leggett1987, Nielsen2010, Palma1996} to electron transfer complexes in quantum chemistry and biology\cite{Mohseni2014, Juzeliunas2000, MerkliJMC2016, Merkli2013}.

For a particle moving in an arbitrary potential $v(x)$  this is the well-known {\it Caldeira-Leggett (CL) model} of quantum Brownian motion \cite{Caldeira1983}. 
Here the inclusion of the counter term \cite{Hanggi2005a} guarantees that the particle dynamics given by the Heisenberg equation of motion for $x$, is determined by the bare potential $v(x)$ and not by a renormalized potential $v(x)-\lambda^2\sum_k|g_k|^2X^2/\omega_k$. Mathematically this is significant since, without the counter term and at sufficiently strong system-reservoir interaction, the energy spectrum of the global system can become unbounded from below leading to a thermodynamically unstable scenario \cite{Ford1988,Ford1997}.
Of particular interest is the quantum harmonic oscillator model for which $v(x)=m\omega_0^2x^2/2$, giving the Hamiltonian \cite{Weiss2008}
\begin{equation} \label{eq:CLHam}
    H_{\rm CL} =\frac{p^2}{2m}+m\omega_0^2 \frac{x^2}{2} 
    +\sum_{k} \frac{p_k^2 
    + \left(m_k \omega_k \, x_k-  \frac{\lambda c_k}{\omega_k} \, x \right)^2}{2m_k}. 
\end{equation}
$\cc_k = -c_k \,\sqrt{\hbar /(m \, m_k\omega_k)}$,
while here $X= x \, \sqrt{m/2}$. 
The Caldeira-Leggett model is a paradigm of open quantum systems \cite{Hu1992,Philbin2016,Funo2018c}, with widespread application to quantum tunnelling \cite{Caldeira1983a} and studies of decoherence and the quantum-classical transition \cite{Hu1992}.

\bigskip

\subsection{Continuum limit for bosonic baths} \label{subsec:contlimit}

For the open system to actually exhibit irreversibility one must take a bath with a continuous spectrum, as we will discuss in detail in Section \ref{sect:RtE} on return to equilibrium. 
In addition there is a practical advantage of taking the continuum limit: it means replacing sums with integrals, and with it converting some intractable summations  into  analytically solvable integrals.

For bosonic baths, one wants to replace the discrete set  $\{\omega_k\}_k$ by a continuum of frequencies, for instance $\omega \in [0, \infty)$. 
There are two common ways of implementing this continuum limit. 
An {\it ad hoc} way 
is to perform the limit in expressions for specific physical quantities (such as time-dependent population probabilities, coherences, etc.) which are obtained from calculations using a discrete mode model, such as \eqref{m1}.
In this approach, the state of the continuous mode model is actually never constructed. The procedure may be rather easily implementable, but it has some disadvantages. For instance, often one has to consider the limits of continuous modes (infinite volume), small/large coupling and large time `simultaneously', and it is not possible to control those limits in this setup. 

The second approach is to immediately construct the continuous mode model \cite{Huttner1992, Anders2020, Cresser2021a, Araki1963, Merkli2020, Konenberg2017} and then analyze the full $\s\b$ statics and dynamics. This allows, in particular, to control perturbation theory for {\it all times}, even $t\rightarrow\infty$. This is done in the quantum resonance theory, which we explain in Section \ref{sec:DaviesArgument1}.
In quantum optics models, the index $k$ labelling the oscillators in \eqref{m1} represents a wave vector in physical space of dimension $d$ (usually, $d=3$) \cite{Nemati2021}.
The continuous mode limit then leads to $k\in{\mathbb R}^d$, and the continuous mode Hamiltonian associated to \eqref{m1},  becomes \cite{Huttner1992,Anders2020}
\begin{equation}\label{m18}
    \HB = \int_{{\mathbb R}^d} \d k  \,  \omega_k  \, a^\dag_k a_k, 
    \quad 
    V_{\s\b} = X \otimes a^\dag(\cc) +{\rm h.c.}, \end{equation}
where  $a^\dag(\cc)=\int_{{\mathbb R}^d} \d k \, \cc_k   \, a^\dag_k$ is the creation operator smoothed out with $\cc_k$ which in this context is sometimes called the {\it form factor}, a square-integrable complex function of $k\in{\mathbb R}^d$. Taking the continuum limit here amounts to taking the quantization volume of the problem to infinity, which also redefines the creation and annihilation operators. In the continuous mode limit, they obey the continuous canonical commutation relations $[a_k , a^\dag_\ell] =\delta(k - \ell)$ where $\delta$ the Dirac  delta-function in $d$ dimensions.  The index $k$ does not need to have a physical meaning though, generally, beyond simply being a continuous index labelling the modes \cite{Huttner1992, Anders2020}. Often it is directly chosen to be  the energy $\omega$ of a mode, which means that   $\HB \propto \int_{0}^{\infty} \d \omega \,  \omega  \, a^\dag_\omega a_\omega$ and   $V_{\s\b} \propto X \otimes \int_{0}^{\infty} \d \omega \, \cc_\omega   \, a^\dag_\omega  +{\rm h.c.}$ 

For a discrete model, an alternative to specifying the coupling constants  $\cc_k$, is to specify the (real) bath spectral density\cite{BrPet2002,Schaller2014,deVega2017},
\begin{equation}\label{eq:specdens}
    J(\omega) := 
    \sum_k |\cc_k|^2 \, \delta(\omega_k-\omega),
\end{equation}
a choice that allows modelling diverse physical situations. For continuous mode environments, the sum is naturally replaced by an integral $\int \d k$.

If $J(\omega)\propto \omega^s$ for small $\omega$, then the spectral density is called  ``Ohmic'' for $s=1$, and ``super-Ohmic'' for $s>1$. 
In what follows we will assume that the spectral density is either Ohmic or super-Ohmic. 
In both cases, the limit $J(\omega)/\omega$ as $\omega\to0$ is finite. This will be important for obtaining finite damping rates in sections \ref{sec:DaviesArgument1} and \ref{Sec:BRME} which turn out proportional to $J(\omega)\coth(\beta\omega/2) \propto J(\omega)/\omega$ at low $\omega$. The sub-Ohmic case  $s<1$ is studied by different theoretical methods see, e.g., Refs.~\cite{Aslangul1987,Grabert1987,Bulla2003,AndersF2007,Winter2009,Alvermann2009,Chin2011,Blunden2017}, and non-Ohmic densities have been found to be relevant in, e.g., opto-mechanical resonator experiments \cite{Groblacher2015}.

\subsection{Exactly solvable $\tmf$ for the Caldeira-Leggett model} \label{subsec:qho}

The properties of the MFG state for the damped quantum harmonic oscillator given by the Caldeira-Leggett Hamiltonian \eqref{eq:CLHam}, for arbitrary coupling strengths $\lambda$, were obtained in Refs.\cite{Grabert1984,Grabert1988,Hanggi2005a}. 
For simplicity, let us formally put here $\lambda=1$ (one can consider $\lambda$ to be included in the coupling coefficients $c_k$). The most explicit results can be obtained in the case of the Drude-Lorentz spectral density: 
\begin{equation} \label{eq:DrudeSpectralDensity}
    J(\omega) =\frac{2 \gamma\omega_{\rm D}}{\pi}
    \frac{\omega\omega_{\rm D}}{\omega^2+\omega_{\rm D}^2},
\end{equation}
in the limit of continuous modes. Here, $\omega_{\rm D}$ is the Drude frequency (which determines the timescale of the bath relaxation) and $\gamma$ is a damping frequency.
Since the CL model is quadratic, the system MFG state will be of Gaussian form and completely determined by the first and second moments \cite{AndersDiplthesis} of the oscillator position and momentum operators: $\langle x\rangle$, $\langle p\rangle$, $\langle x^2\rangle$, $\langle p^2\rangle$, and $\langle px\rangle$. The first moments trivially vanish, while the second moments are given in terms of the partition function $ Z(\omega_0,\gamma)$ at inverse temperature $\beta$,
\begin{equation}
    Z(\omega_0,\gamma)=\frac{\beta\omega}{4\pi^2}
    \frac{\Gamma(\mu_1/\nu)
    \Gamma(\mu_2/\nu)\Gamma(\mu_3/\nu)}
    {\Gamma(\omega_{\rm D})},
\end{equation}
where $\Gamma(z)$ denotes the gamma function and $\nu=2\pi/\beta$ is the first {\it Matsubara frequency}. 
The functions $\mu_j (\omega_0, \gamma)$, for $j=1,2,3$, denote the roots of the cubic polynomial 
\begin{equation}
    \mu^3-\omega_{\rm D}\mu^2+(\omega_0^2+\gamma\omega_{\rm D})\mu-\omega_{\rm D}\omega_0^2.
\end{equation}
With these expressions, the second moments given in unit-free form ($\tilde{x}$ and $\tilde{p}$ see \eqref{eq:CLHam}), and including $\hbar$ explicitly, are \cite{Grabert1984}   
\begin{eqnarray} \label{eq:qhomoments}
    \langle \tilde{x}^2\rangle 
    &=&\frac{m \omega_0}{\hbar} \langle x^2\rangle 
    =-  \frac{1}{\beta  \hbar}     \frac{\partial\ln Z(\omega_0,\gamma)}{\partial\omega_0}, \nonumber \\
    \langle \tilde{p}^2\rangle 
    &=& \frac{1}{m \omega_0 \hbar} \langle p^2\rangle 
    = \langle \tilde{x}^2\rangle      
    - \frac{2\gamma}{\beta \hbar \omega_0 } \frac{\partial\ln Z(\omega_0,\gamma)}{\partial\gamma},\\ 
    \langle \tilde{p} \tilde{x}\rangle 
    &=& \frac{1}{\hbar}   \langle px\rangle 
    =- \frac{i}{2}. \nonumber
\end{eqnarray}
Ref.\cite{Grabert1984} implies the MFG state as the Gibbs state of an effective harmonic oscillator Hamiltonian (the HMF) $\HS^{\rm eff} = {p^2 / (2 m(\lambda))} + {m(\lambda) \omega_0^2(\lambda) x^2 / 2}$. They formally establish the mass $m(\lambda)$ and frequency $\omega_0(\lambda)$ of the oscillator which are rescaled from their bare counterparts due to the interaction with the bath. Alternatively, the harmonic oscillator's MFG state can also be given in Gaussian integral form  \cite{AndersDiplthesis}, expanded in the Weyl-basis, as
\begin{eqnarray} \label{eq:harmonicoscitmf}
    \tmf=\frac1{2\pi}\int_{\mathbb R^2}\d\xi_1 \d\xi_2 \, \, 
    e^{- \xi_1^2 \, \langle \tilde{p}^2\rangle /2} \, e^{ - \xi^2_2 \, \langle \tilde{x}^2\rangle/2 }
    \, \, e^{i(\xi_1 \, \tilde{p} -\xi_2 \, \tilde{x})}. \quad
\end{eqnarray}
It is an open question to show if  this expression indeed reduces to the Gibbs state of the effective Hamiltonian given in \cite{Grabert1984}.

\smallskip

Another method of obtaining the oscillator moments, cf.  \eqref{eq:qhomoments}, is via the closed expressions for the oscillator correlation functions in the MFG state.  These have been derived \cite{Philbin2012,Philbin2016} using Heisenberg equations of motion for an arbitrary spectral density $J(\omega)$, e.g.
\begin{multline} \label{eq:Philbinx^2}
    \langle x(t)x(t')+x(t')x(t)\rangle\\=
    \frac{1}{\pi}    \int_0^\infty \d\omega\,
    \cos[\omega(t-t')] \, 
    \coth\left(\tfrac{\beta\omega}2\right) \, 
    {\rm Im}\, {\mathcal G}(\omega),
\end{multline}
where the mass was set to $m=1$ and $\hbar =1$ again. 
Here 
 $   {\mathcal G}(\omega)=-1/(\omega^2-\omega_0^2[1-\chi(\omega)])$
is a Green's function and $\chi(\omega)$ is a complex susceptibility given by
\begin{equation}\label{eq:PhilbinChi}
    \omega_0^2 \, \chi(\omega)=\int_0^\infty
    \d\xi\,\frac{\omega J(\omega)}{\xi^2-\omega^2}
    +\frac{i\pi J(\omega)}2,
\end{equation}
where the integral is understood as the principal part integral. 
Setting $t=t'$ makes the correlation function  \eqref{eq:Philbinx^2} time-independent, and it should then correspond to the first line in \eqref{eq:qhomoments}. 
This route allows to give analytic expressions for thermal energies \cite{Philbin2016} of damped quantum and classical harmonic oscillators as a functional of $\chi(\omega)$ and inverse temperature $\beta$.

\subsection{Expansion of $\tmf$ for small coupling constant $\lambda$} 
\label{subsec:staticweak}
\label{subsec:weakenough}

We are now interested in expressions for the MFG state in the weak coupling limit, when $\lambda$ in \eqref{eq:Htot} is assumed to be small, so that an expansion of $\tmf$ can be considered. Such expansions are based on the Kubo identity \cite{Toda2012} and are sometimes referred to as ``canonical perturbation theory''\cite{Subasi2012}, equivalent to a standard time dependent perturbation expansion with $t$ replaced by $-i\beta$. In the mathematical literature, the expansion in $\lambda$ is known as the perturbation theory of KMS (Kubo-Martin-Schwinger) states \cite{BratteliRobinson1981}. For a bosonic bath and a linear $\s\b$ coupling as in \eqref{m18}, 
one can consider the expansion
\begin{equation}\label{eq:tauMFseries}
    \tau_{\rm MF}=\tau+\lambda^2\tau_{\rm MF}^{(2)}
    +\lambda^4\tau_{\rm MF}^{(4)}+\ldots .
\end{equation}
Only even terms in $\lambda$ are present because thermal equilibrium averages of products with any odd number of bath creation and annihilation operators vanish. 
Such expansions have been provided in Refs.~\cite{Geva2000,Mori2008,Thingna2012,Subasi2012,Purkayastha2020,Cresser2021a} to second order in $\lambda$. In Ref.~\cite{Thingna2012}, the correction term $\tau_{\rm MF}^{(2)}$ was obtained for the Caldeira-Leggett model (with an arbitrary potential $v(x)$), and in Ref.~\cite{Purkayastha2020} for the spin-boson model.

\smallskip

Ref.~\cite{Cresser2021a} considers an arbitrary system $\s$ with system Hamiltonian $\HS$ with discrete spectrum,  coupled to a continuous bosonic bath via an arbitrary Hermitian system operator $X$.
The Hamiltonian, including the counter term ({\em c.f.} Eqs.~ \eqref{eq:Hwcounter} and \eqref{m18}) is given by 
\begin{equation}\label{eq:totalHamiltonian}
	\Htot =\HS + \frac12 \int_{0}^\infty \d\omega \, \left[ p^2_\omega + \left(\omega \, q_\omega+\lambda {\sqrt{\frac{2 \J(\omega)}{\omega}}} \, X \right)^2\right].
\end{equation}
Note that here a spectral density $\J(\omega)$ is used immediately, without specifying any coupling constants $c_k$. 
Here $\left[q_\omega, p_{\omega'}\right]=i \, \delta(\omega-\omega')$ are the commutation relations for the bath position and momentum operators.
Using perturbative methods and evaluating the bath trace  in \eqref{eq:MFGibbs} explicitly,  the dominant  correction $\tmf^{(2)}$ for an arbitrary system was found to be 
\begin{equation} \label{eq:NormalizedMFGSmain}
    \begin{split}
    \tmf^{(2)}= & \beta\sum_m \tau
    \left(
    X_mX_m^\dagger-\trS[ \tau \, X_m \, X_m^\dagger]
    \right)
    \D_\beta(\omega_m)
    \\
    &+ \sum_m[X_m^\dagger, \tau  \, X_m]
    \frac{d \D_\beta(\omega_m)}{d\omega_m}
    \\
    &+ \sum_{m\neq n}
    \left(
    [X_n,X_m^\dagger \, \tau]+\text{h.c.}
    \right)
    \frac{\D_\beta(\omega_m)}{\omega_n-\omega_m}.
    \end{split}
\end{equation}
Here, the decomposition of the Hermitian system operator $X$ into a sum of energy eigenoperators $X_m$  is used, where the $X_m$ are defined by
\begin{equation}\label{m16}
	\left[H_\s,X_m\right]=\omega_m \, X_m;\quad X_{-m}=X_m^\dagger,\quad \omega_m=-\omega_{-m},
\end{equation}
with $\omega_m$ the Bohr frequencies of the system (energy differences of $\HS$). 
Furthermore, the function $\D_\beta(\omega_m)$ is defined as
\begin{equation}\label{eq:Dbeta}
    \D_\beta(\omega_m)=
    \int_0^\infty \d \omega \,  \J(\omega) \, 
    \left(
    \frac{\omega_m\coth(\beta\omega/2)+\omega}
    {\omega^2-\omega_m^2}
    -\frac1\omega
    \right),
\end{equation}
where the integral is understood as a principal part integral.
Expression \eqref{eq:NormalizedMFGSmain} evidences the appearance of coherences in the $\HS$ basis in the system's equilibrium state $\tmf$. Coherences are often considered a quantum `resource' \cite{Streltsov2017}, and beyond their significance in quantum thermodynamics \cite{Uzdin2015,Kammerlander2016,Francica2020,Messinger2020,Purkayastha2020,Hammam2021}, play an important role in some biological processes \cite{Lloyd2011,Lambert2013,Jeske2015,Dodin2016,Dodin2018}.
\smallskip 

One can now  also quantify what is ``weak enough'' for the  weak coupling limit and expression \eqref{eq:NormalizedMFGSmain}  to be valid.
Beyond the loose requirement that $\lambda$ ought to be ``small'', one finds (by comparing perturbative orders) that $\lambda$ has to obey the inequality \cite{Cresser2021a}  
\begin{equation}\label{eq:binomialApproxConditionmain}
	|\lambda| \ll \frac{1}{\sqrt{| \beta\sum_{m}\trS\left[\tau \, X_m X_m^\dagger\right] \D_{\beta}(\omega_m)  | }}.
\end{equation}
This condition gives a well-quantified limit for $\lambda$ being in the weak coupling regime at a given $\beta$. Note that the range of $\lambda$ for which the weak coupling regime and hence \eqref{eq:NormalizedMFGSmain} is applicable changes as a function of temperature, with larger temperature generally allowing larger $\lambda$.
\subsection{Ultrastrong coupling for general system and bosonic bath}  \label{subsec:staticultrastrong}

One can also consider the opposite limit, when the  coupling is much stronger than other energy scales of the system, i.e. the ``ultrastrong'' coupling limit $\lambda \to \infty$, see Fig.~\ref{fig:weakvsstrong} and Refs.~\cite{Lambert2019,Yu2021,Goyal2019,Acharyya2020,Pilar2020}. Here it is also possible\cite{Cresser2021a} to find an explicit expression for $\tmf$ for a general system. This is done for the case that it couples to a bosonic bath as in \eqref{eq:totalHamiltonian}, with a single system interaction operator $X$ with non-degenerate spectrum (extensions to degenerate situations should be straightforward),
\begin{equation}\label{eq:Xspectr}
    X=\sum_n x_n P_n,    
\end{equation}
where $x_n$ are real numbers and  $P_n$ are orthogonal projectors of rank one.
By expanding the global Gibbs state $\tauglobal$ in orders of $1/\lambda$ and explicitly integrating out the bath oscillators, the mean force Gibbs state simplifies to\cite{Cresser2021a}
\begin{equation}\label{eq:MFGultrastrong}
    \tmf =\frac{e^{-\beta \sum_n P_n \, \HS \,  P_n}}{ \trS[e^{-\beta \sum_m P_m \, \HS \, P_m}]}.
\end{equation}
This is a surprisingly neat form for the open system equilibrium state at ultrastrong coupling. It implies that, in this limit, the equilibrium state of the system becomes diagonal in the basis of the system's coupling operator $X$. In the context of  measurement and decoherence theory it is referred to as the `pointer basis' \cite{Zurek1981,Zurek2003,Eisert2004,Goyal2019,Orman2020}. For $[\HS, X] \neq 0$, an immediate consequence is that $\tmf$ will maintain coherences in the $\HS$ energy basis $|e_m\rangle$, i.e. $\langle e_m |\tmf|e_m \rangle \neq 0$ for some $m$. Corrections to Eq.~(\ref{eq:MFGultrastrong}) with respect to $\lambda^{-1}$ have been obtained in Ref.~\cite{Latune2021}.

It is worthwhile to build bridges between these results and discussions about localized and delocalized excitations in the theory of excitation energy transfer in biological photosynthetic complexes\cite{Huelga2013, Fassioli2013,Jang2018,Mohseni2014}. Due to the dipole interaction between the chromophore molecules, the eigenstates of the system Hamiltonian are superpositions of {\it local excitations} and describe the so called delocalized excitons. The $X$ operator (or more generally the $X^{(j)}$ in Eq.~\eqref{vint}) is diagonal in the local excitation basis. So, the local excitation basis corresponds to the pointer basis.

In this context, ``weak coupling'' theory describes the relaxation in the basis of delocalized excitons. Non-negligible coupling to the phonon bath leads to the relaxation not in the exciton basis, but in a more localized basis\cite{Fassioli2013}. 
In the ultrastrong coupling limit, which corresponds to F\"orster theory of excitation energy transfer,\cite{Mohseni2014} the relaxation occurs in the local excitation (pointer) basis. This is exactly the regime described by the mean force Gibbs state \eqref{eq:MFGultrastrong}. 
The dynamical aspects of the F\"orster regime of excitation energy transfer and the ultrastrong coupling regime for a general open quantum system will be discussed in Sec.~\ref{sec:ultrastrongdyn}.

\subsection{Intermediate coupling: Polaron transformation} \label{subsec:staticpolaron}
\label{sec:polaronstatic}

Under certain conditions, the ultrastrong and intermediate coupling regime, see Fig.~\ref{fig:weakvsstrong}b), can be treated  with the so-called {\it polaron transformation}.
Originally, a polaron is a quantum quasiparticle in a solid material consisting of an electron and a field of elastic deformations of the crystal lattice (a phonon cloud)\cite{Landau1933}. In a more general context of open quantum systems, a polaron is a state of the system ``dressed'' by the bath excitations. 
Mathematically, the polaron transformation is a certain unitary transformation acting on $\s\b$, which mixes the system and bath DoFs\cite{Holstein1959,Rachkovsky1973,Abram1975,Silbey1984,Harris1985}.  
The benefit of the polaron transformation is that one can apply weak coupling perturbation theory for the redefined system and bath; see also Section \ref{sec:polarondyn}.

As an illustration of this formalism, we consider the spin-boson model \cite{Lee2012jcp,Lee2012pre,Xu2016, Leggett1987}. The total  Hamiltonian \eqref{eq:totalHamiltonian} here contains \begin{equation}\label{eq:spinHS}
    \HS=\frac\varepsilon2\sigma_z+\frac\Delta2\sigma_x,
\end{equation}
and the system coupling operator is $X=\sigma_z$, where $\sigma_{x,y,z}$ are the usual Pauli matrices. Thus the pointer basis is the $\sigma_z$-basis.
Then, the polaron transformation is given by the unitary transformation 
\begin{equation}
     U=\exp(-i\sigma_z\otimes \R)  
     \quad \mbox{with} \quad  
    \hat{R} =   \int_0^\infty \d\omega \, \sqrt{\frac{2 \J(\omega)}{\omega}} \frac{p_\omega}{\omega}.
\end{equation}
The polaron-transformed Hamiltonian  (indicated with a tilde) is
$ \tilde H_{\rm tot}=U\Htot U^\dagger=\tilde H_\s+H_B+ \tilde V_{\s\b}$,
where $\tilde \HS=\frac\varepsilon2\sigma_z+
    \kappa \,  \frac{\Delta}2\sigma_x$  
with 
\begin{equation}
    \kappa :=\trB[\tau_\b \, \cos (2 \R)]=
    \exp\left[-2 \lambda^2 \int_0^\infty \d \omega \, \frac{\J(\omega)}{\omega^2}
    \coth\left(\tfrac{\beta\omega}{2}\right)
    \right].
\end{equation}
The bath part remains unchanged, $\HB =\int_0^{\infty} \, \,d\omega \, \omega \, a_\omega^\dagger a_\omega$, 
and the interaction becomes
\begin{eqnarray}
    \tilde V_{\s\b} &=& 
    \sigma_x\otimes B_x+\sigma_y\otimes B_y, \nonumber \\
    B_x&=&\frac{\Delta}2(\cos (2 \R) -\kappa) \quad \mbox{and} \quad
    B_y=\frac{\Delta}2 \, \sin (2 \R) ,
\end{eqnarray}
where $\lambda$ has now moved inside $\tilde{V}_{\s\b}$.
When the integral in the definition of $\kappa$ converges, it turns out that $\trB[\tau_\b \, e^{\pm i 2 \R }]=\kappa$. But convergence only happens for a subclass of super-Ohmic spectral densities. For example, the integral converges whenever,  for small $\omega$  and considering strictly positive temperatures,  $\J(\omega)$ is proportional to $\omega^3$, but it diverges whenever $\J(\omega)$ is proportional to $\omega^s$ for $s\leq2$. This is a restriction of the polaron transformation method.

The factor $\kappa$ represents the above mentioned ``phonon cloud'', while the $B_x$ and $B_y$ operators represent fluctuations around this cloud. One may hope that these fluctuations are not large and $\tilde V_{\s\b}$ can be treated perturbatively. Thus, the benefit of the polaron transformation is that one can now apply the weak coupling perturbation theory for the rotated $\s\b$ complex. 
However, strictly speaking, the conjecture of applicability of the weak coupling theory to the polaron-transformed Hamiltonian is justified only in two opposite limits: (i) the weak system-bath limit (small $ \lambda^2 \J(\omega)$), where the polaron transformation is trivially reduced to the identity transformation, 
and (ii) the limit of small $\Delta$ (weak tunneling limit)\cite{Xu2016,Kolli2011}.
Since  $|\kappa\Delta|<|\Delta|$  the eigenvectors of $\tilde H_\s$ are more `localized' superpositions of the pointer basis vectors, than the eigenvectors of $\HS$. Such localization due to non-negligible system-bath interaction was mentioned at the end of Sec.~\ref{subsec:staticultrastrong}.

Now we can consider the total Gibbs state in the polaron frame $\tilde\tau_{\s\b} = U \, \tauglobal \, U^\dag  \propto e^{-\beta\tilde H_{\rm tot}}$. One can show that the diagonal part (in the pointer basis) of the desired MFG state $\tmf$  formally coincides with the diagonal part of  the reduced system state $\tilde\eta=\trB[\tilde\tau_{\s\b}]$, i.e. $\tau_{{\rm MF},jj} =\tilde\eta_{jj}$ for $j=1,2$.
Approximate expressions for $\tilde{\eta}$ were obtained \cite{Lee2012jcp,Lee2012pre,Xu2016} using second-order perturbation theory with respect to $\tilde V_{\s\b}$, 
\begin{equation} \label{eq:polaronexpansion}
    \tilde{\eta} \approx \tilde{\eta}^{(0)} + \tilde{\eta}^{(2)} + {\cal O}(\tilde V_{\s\b}^4),
\end{equation}
where $ \tilde{\eta}^{(0)}  
={e^{-\beta\tilde H_\s} / \tilde{Z}}$ with $ \tilde{Z} = \trS [e^{-\beta\tilde H_\s}]$ is the Gibbs state corresponding to the system Hamiltonian in polaron frame. 
The next term in \eqref{eq:polaronexpansion} is
\begin{eqnarray}
    \tilde{\eta}^{(2)} 
    &=&\frac{A}{\tilde Z}
    -\frac{\trS[A]}{\tilde Z} \, \tilde{\eta}^{(0)},
\end{eqnarray}
where
\begin{gather*}
        A=\sum_{m,n=x,y}
        \int_0^\beta \d\beta'\int_0^{\beta'}\d\beta''
        G_{mn}(\beta'-\beta'')
        e^{-\beta\tilde H_\s}\sigma_m(\beta')\sigma_n(\beta''),
\end{gather*}
and $G_{mn}(\beta)=\tr[B_m(\beta)B_n\tau_\b]$ are the imaginary-time bath correlation functions. The operators in imaginary time, such as $\sigma_m(\beta)$ and $B_m(\beta)$, are defined as $O(\beta)=e^{\beta(\tilde H_\s+H_\b)}Oe^{-\beta(\tilde H_\s+H_\b)}$. These expressions can be made more explicit using the methods of Ref.~\cite{Cresser2021a} presented in Sec.~\ref{subsec:staticweak}.

But to determine the off-diagonal element $\tau_{{\rm MF},12}$ (in pointer basis),  $\tilde\tau_{\rm MF}$ does not suffice -- the bath DoF of the total polaron-transformed Gibbs state are also required. 
Up to the first order in $\tilde V_{\s\b}$ it is found \cite{Lee2012pre,Xu2016} to be 
\begin{equation}\label{eq:polaronmfgcoherence}
    \tau_{{\rm MF},12} \approx  
    -\tfrac{\kappa^2\Delta}{2\Lambda}\tanh(\tfrac{\beta\Lambda}{2})
    -\sum_{m=x,y} \int_0^\beta\, \d\beta' \, S_m(\beta') \, K_m(\beta'),
\end{equation}
where $\Lambda=\sqrt{\varepsilon^2+(\kappa\Delta)^2}$, $S_m(\beta')=\trS[\sigma_m(\beta') \, \sigma_- \, \tilde\tau]$, and $K_m(\beta')=\trB[V_m(\beta') \, \cos (2 \R)  \, \tau_\b]$,  where the functions $S_m$ and $K_m$ can be explicitly evaluated and $\sigma_{-}=(\sigma_x-i\sigma_y)/2$.

In Refs.~\cite{Lee2012jcp,Lee2012pre,Xu2016}, the following (super-Ohmic) spectral density is considered:
\begin{equation}
 \J(\omega)=\frac{\gamma}{2}
 \frac{\omega^3}{\omega^3_{\rm c}}e^{-\omega/\omega_{\rm c}},
\end{equation}
where $\gamma$ (or, more precisely, $\lambda^2\gamma$) determines the system-bath coupling strength, while $\omega_{\rm c}$ is the cutoff frequency and determines the rate of relaxation of the bath correlation functions in time. The above expressions for the elements of $\tmf$ are compared with  numerically exact simulations. It turns out that the approximation works well for the cases of fast bath $\omega_{\rm c}>\Delta$ and  ultrastrong coupling (large $\lambda$).
It remains an open question to simplify expressions \eqref{eq:polaronexpansion} for  $\tmf$  to a form similar to \eqref{eq:NormalizedMFGSmain}.

Finally, the so-called variational (partial) polaron transformation can be used to enlarge the range of applicability of this approach. 
In particular, the variational polaron transformation allows one to overcome the assumption of the super-Ohmic spectral density. Numerical calculations of the equilibrium state and equilibrium physical observables for the Ohmic spectral baths using the variational polaron approach are presented in  \cite{Lee2012jcp,Lee2012pre,Xu2016,Popovic2021}.

\subsection{High-temperature expansion}

An approximate expression for $\tmf$ in the high temperature limit can be obtained \cite{Gelzinis2020} by expanding  in powers of inverse temperature $\beta$.

In the notations introduced above, the model considered in Ref.~\cite{Gelzinis2020} can be formulated as follows: a multi-state system with orthonormal basis $\{\ket{n}\}$ interacts, via  $X_n=\ket{n}\bra{n}$, with several baths $n=1, \ldots, N$ that all have the same temperature. Instead of  \eqref{eq:totalHamiltonian}, the Hamiltonian is
\begin{equation} \label{eq:multilevelH}
	\Htot =\HS + \frac12\sum_{n=1}^N \int_{0}^\infty \d\omega \, \left[ p^2_{\omega, \, n} + \left(\omega \, q_{\omega, \, n}+\lambda {\sqrt{\tfrac{2 \J_n(\omega)}{\omega}}} \, X_n \right)^2\right],
\end{equation}
for independent baths
\begin{equation*}
    [q_{\omega, \, n},p_{\omega', \, m}]=
i\delta(\omega-\omega') \, \delta_{nm}.
\end{equation*}
The system Hamiltonian can be decomposed as $\HS=H_\epsilon+H_J$ where $H_\epsilon$ is the part that is diagonal in the $\ket{n}$ basis and  $H_J$ contains all off-diagonal contributions (cf.\ \eqref{eq:PointerBasisSeparation}). 
Expanding \eqref{eq:MFGibbs} for the Hamiltonian \eqref{eq:multilevelH} to second order in $\beta$, evaluating the partial traces, and then  re-summing into an exponential form, gives a  MFG state  \cite{Gelzinis2020} proportional to
\begin{equation} \label{eq:highTMF}
    \tau_\text{MF}\propto\exp\left[-\beta(H_\epsilon+e^{-\frac{1}{6}\beta\Lambda} \, H_J \, e^{-\frac{1}{6}\beta\Lambda})\right].
\end{equation}
Here $\Lambda=\lambda^2\sum_{n=1}^N
\int_0^\infty
d\omega \, 
\frac{J_n(\omega)}{\omega} \, X_n
\equiv \sum_{n=1}^N \ell_n \, X_n$
is an operator-valued reorganization term.
Result \eqref{eq:highTMF} implies that the inter-state coupling constants $J_{mn}$ contained in $H_J$, describing hopping between states $\ket{m}$ and $\ket{n}$, get rescaled by the bath interaction into effective coupling constants. The rescaling itself is temperature dependent and vanishes for $\beta \to 0$. For a dimer system with $\ell_1=\ell_2=\ell$, expression \eqref{eq:highTMF} is found to be accurate  \cite{Gelzinis2020} for temperatures satisfying $\ell\beta\lesssim2$. We emphasise that generally Eq.~\eqref{eq:highTMF} is valid even at intermediate system-bath coupling strengths as long as the temperature is large enough.

\subsection{Adequacy of bosonic bath model}
\label{Sec:OnFreeBath}

The bosonic bath model considered from section \ref{subsec:discbosonic} onwards is very widely used in the theory of open quantum systems and quantum thermodynamics, though fermionic and spin bath models are also actively studied \cite{Davies1974,Segal2014, Jing2018,Prokofev2000,Sharma2015,Hamdouni2007,Breuer2004a,Breuer2007}.
In addition to it being bosonic, the coupling to the system is assumed to be linear in creation and annihilation operators. In this case, the bath DoFs can, without loss of generality, be assumed to be non-interacting, as re-diagonalization can always bring it into such a normal mode form.
Nevertheless, the resulting bath model is a very special case, and we here discuss its range of applicability.

One can distinguish three levels of validity of this model. The only case where this model is exact is for a bath consisting of photons (electromagnetic field) \cite{CohenTann1992,Carmichael1993,Carmichael1999,Agarwal2012,Cottet2017,MayKuhn,Mukamel}. On the second level, the bosonic bath model is used as an approximation of the real physics. 
For example, in solid-state physics and chemistry, e.g. for modelling charge and energy transfer, the bath consists of phonons or vibrational modes that describe oscillatory degrees of freedom of nuclei. 
If the magnitudes of these oscillations are not too large, the harmonic approximation can be used \cite{MayKuhn,Valkunas2013} implying linear coupling to the system (electronic degrees of freedom). 
This approximation is used for various system-bath (electron-phonon) coupling regimes \cite{MayKuhn}. 
Even at strong coupling, it is often a reasonable assumption that the nuclei oscillations around the equilibria are small enough to warrant the harmonic approximation (though the equilibria themselves can be significantly shifted due to strong coupling). \cite{Caldeira1983a}

The third level is the use of this model as a phenomenological model, which need not directly represent the real physics of the bath. Namely, let the bath be a complex system of many interacting particles that cannot be reduced to a set of harmonic oscillators. But the details of the bath dynamics are not that important for the reduced dynamics of the system - only some aggregated properties of the bath dynamics, such as correlation functions, are required. 
Besides, Gibbs states of the bosonic bath are Gaussian, implying that all correlation functions can be expressed in terms of just the second-order correlation functions. 
Such Gaussian property is likely to emerge also for rather general baths containing a large number of particles,
as a consequence of the central limit theorem. 
So, there is hope that a bosonic bath with the same second-order correlation functions as that of a real bath may serve as a phenomenological model of the real bath, at least for qualitative analysis.

\section{Return to Equilibrium} \label{sec:dynamicsA}

Return to Equilibrium (RtE) is a  basic and intuitive phenomenon, saying that initial states which do not deviate much from the equilibrium state, will converge to the equilibrium state in the long time limit. As an analogy, a ripple created at some point on the surface of a still lake  (equilibrium)  will propagate away. Eventually the lake's surface will return to be still. For this to happen, it seems clear that the total system under consideration has to be infinitely large to avoid recurrences for all times, and that the dynamics has to be dissipative in the sense that it propagates local disturbances away to infinity. Furthermore, the  perturbation must be `small', for instance localized in space (if initial ripples are created everywhere in space then at any fixed point, the surface will not remain still, even for large times, as ripples keep arriving from far away positions). 
In the following section we formalise these intuitive notions in mathematical terms.

\subsection{Continuous spectrum and the emergence of irreversibility} \label{sec:continmodes}

In the static approach, see Sec.~\ref{subsec:MFGstate}, we have assumed a super-bath to justify the system+bath Gibbs state $\tauglobal$ as the equilibrium state. This immediately implied that the MFG state $\tmf$ is the equilibrium state for the system alone. 
Now we abandon the super-bath 
and adopt the point-of-view that $\s\b$ together forms a {\it closed} system complex. In the following, we will identify system and bath properties which lead to "self-thermalization", that is, the convergence towards the global Gibbs state $\tau_{\s\b}$ (return to equilibrium).

The Hamiltonian $\Htot$ of the total, closed $\s\b$ complex determines the dynamics in time $t$ from an initial $\s\b$ state $\rho_{\s\b} (0)$  according to the Schr\"odinger equation,
\begin{equation}\label{m2}
    \rho_{\s\b}(t) 
    =    e^{-i t \Htot} \, \rho_{\s\b}(0)\,  e^{i t \Htot}.
\end{equation}
$\Htot$ also determines the global Gibbs state $\tauglobal$ at inverse temperature $\beta$, see Eq.~\eqref{eq:tausb}.
We now discuss two mathematical intricacies relating to equilibration towards $\tauglobal$, issues that have been kept under the carpet in the Statics section~\ref{sec:statics}. 

\smallskip

There are formal definitions of irreversibility\cite{Schmitt2018} 
-- here we mean by it, somewhat intuitively, that averages of suitable observables approach constant values in the limit of large times. The first point to recall is that a {\it closed} system exhibits truly irreversible dynamics  {\it only} if its Hamiltonian $H$ has a continuous energy spectrum \cite{Bocchieri1957}. Indeed, if the energies are discrete, $E_1,E_2,\ldots \in \mathbb R$, with corresponding eigenstates $|\psi_j\rangle$, then the evolution is $e^{- i t H} = \sum_j e^{-i t E_j} |\psi_j\rangle\langle\psi_j|$. This shows that the dynamics is simply oscillating for all times.  

The connection between irreversible dynamics and continuity of modes (energy spectrum) can be illustrated on a bath consisting of non-interacting particles as follows.  Consider a closed system of $n$ particles in a region $V\subset {\mathbb R}^3$ with  Hamiltonian (kinetic energy) 
\begin{equation}
    \label{eq:nparthamilt}
H_V=-\sum_{j=1}^n \frac{\Delta_j}{2m_j},
\end{equation}
where $\Delta_j=\partial_{x_j}^2$ is the Laplacian. For {\it finite} $V$, the energies of $H_V$ are  discrete and the dynamics generated by $H_V$ is quasi-periodic (a sum of oscillating terms). Particles are confined to $V$ and boundary effects cause recurrence. 
However, for $V={\mathbb R}^3$, the spectrum of $H_V$ becomes continuous (equal to $[0,\infty)$). The dynamics is now irreversible in the following sense: Given an arbitrary {\it finite} region of observation $R\subset{\mathbb R}^3$, the probability of finding any of the particles inside $R$ converges to zero in the limit of large times (the particles travel to infinity). Of course, even in the infinite volume situation, the probability of finding the particles in all of space ${\mathbb R}^3$ equals $100\%$ at all times. This is simply a consequence of the global unitarity of $e^{-i t H_V}$. However, all observations made in any finite volume (such as a laboratory), show dynamical irreversibility.
\footnote{
Another way of understanding the connection between continuous modes and irreversibility is provided by results on weak convergence of solutions of the Liouville equation (in classical mechanics). Though the dynamics of a classical particle in a bounded domain is reversible and recurrent, if we consider the corresponding Liouville equation with a continuous initial state (density function), then a weak convergence to a stationary state on long times (both positive and infinite long times) can be proved, which was originally observed by Poincar\'{e}\cite{Poincare1906} and developed in Refs.~\cite{Kozlov2002book,Kozlov2002,Kozlov2003,Kozlov2007}}

\smallskip

The second point to highlight is that in writing \eqref{eq:tausb} one implicitly assumes that the matrix $e^{-\beta \Htot}$ is trace-class (meaning that its trace is finite).
However, for Hamiltonians $\Htot$ which have  continuous spectrum,  we always have 
\footnote{If a non-negative operator has finite trace, then the operator is compact, which in turn implies that its spectrum consists of discrete eigenvalues only. See for instance \cite{deGosson2011}.}
$\tr[e^{-\beta \Htot}]=\infty$,
and so the equilibrium state cannot be expressed by \eqref{eq:tausb}. 
In this situation, one must in fact use a limiting procedure to mathematically define the equilibrium state, as we illustrate in the next subsection (see also Section 4 of  Ref. \cite{Merkli2006} and Ref. \cite{BratteliRobinson1981}).

\smallskip

From now on we consider the bath to be a very large and complex environment, with a Hamiltonian $\HB$ having a continuum of energies. In contrast, the system is ``small and simple'', typically having a Hamiltonian $\HS$ with discrete ({\em i.e.} not continuous) spectrum.
The total interacting  Hamiltonian $\Htot$ for the $\s\b$ complex, see \eqref{eq:Htot}, then inherits the property of continuous energies. In a sense, if you add to a very complex physical system (the bath $\b$) some more degrees of freedom (by coupling it to a small system $\s$), then the overall characteristics of the energy spectrum does not change (continuous spectrum stays continuous spectrum under such a coupling). 

Despite our above convention of using the term ``bath'' in the case of continuous modes (infinite volume), we sometimes want to discuss the situation of large but finite `baths', in which case we use the term ``finite bath''.

Before discussing the continuum limit of $e^{-\beta \Htot}$ in Sec.~\ref{sect:infvolequilibrium}, we first comment on the decription of (apparent) irreversibility  for non-continuous systems.

\subsection{Finite baths \& effective dimension} 
\label{subsect:effdim}

While any physical lab environment is large - but finite - taking the continuum limit is a meaningful approximation of most real situations, where a multitude of uncontrolled and spatially far extended bath modes may interact with a system of interest. The question in what way a very large, but finite environment can describe thermalization or, more generally, equilibration (convergence to an equilibrium state which is not necessarily thermal), is addressed in Refs.\cite{Reimann2008,Linden2009}

The authors consider a generic class of $\Htot$, with non-degenerate eigenvalues and non-degenerate Bohr frequencies different from zero (Bohr frequencies are differences between the eigenvalues). 
They show that the average magnitude of fluctuations in time, around the equilibrium state (generally dependent on the initial state), is proportional to $1/\sqrt{d_{\rm eff}}$, where the effective dimension is defined by
\begin{equation}\label{eq:deff}
    d_{\rm eff}=\frac{1}{\tr[\overline{\rho}_{\s\b}^2]},
\end{equation}
in which $\overline{\rho}_{\s\b}$ is the time-averaged state,
\begin{eqnarray}
    \overline{\rho}_{\s\b}
    &:=& \lim_{\tf \to\infty} \frac{1}{\tf} \int_0^{\tf} \d s  \, e^{-i s \Htot} \, \rho_{\s\b}(0) \, e^{i s \Htot}  \nonumber
\\
    &=& \sum_k \langle E_k|\rho_{\s\b}(0)|E_k\rangle\, |E_k\rangle\langle E_k|,
\end{eqnarray}
where $|E_k\rangle$ are the eigenvectors of $\Htot$, see \cite{Reimann2008,Linden2009}. The quantity $\tr[\overline{\rho}_{\s\b}^2]$ is equal to the time average of the Loschmidt echo, or the survival probability \cite{VenutiZanardi2015} of the initial state, i.e.
\begin{equation}
    \tr[\overline{\rho}_{\s\b}^2]
    =\lim_{\tf\to\infty}\frac{1}{\tf} \int_0^{\tf} \d s \, \tr[e^{-i s \Htot} \, \rho_{\s\b}(0) \, e^{i s \Htot} \,  \rho_{\s\b}(0) ].
\end{equation}
The recurrence time grows exponentially with $d_{\rm eff}$, see Ref.~\cite{Venuti2015} 
There are estimates for interacting many-body systems\cite{VenutiZanardi2015} indicating that $d_{\rm eff}$ is exponentially large in the joint system plus bath size.
Indeed, strong  evidence exists that $d_{\rm eff}$ is exponentially large for almost any wavefunction. \cite{Bengtsson2017,Popescu2006} 
One may conjecture that $d_{\rm eff}$ increases indefinitely with the number of modes; but so far, a rigorous proof  of this for the bosonic bath model described above has not been achieved.

As the recurrence time increases with the bath size, taking the infinite volume, or continuous modes limit, corresponds to setting the recurrence time to infinity. This is physically not always realistic, and when it isn't, then one has to study the system and bath dynamics for finite baths, which is an intricate task, as it necessitates simultaneously the analysis of the non-trivial (non-constant) bath dynamics. It has been observed that for certain models, the dynamics converges numerically to a stationary regime on rather fast time scales, even if the bath consists of only a relatively small number of degrees of freedom.\cite{ASP2021,ASP2021(2),Esposito2003,Esposito2007,Pozas2018,Lotshaw2019}

\subsection{Constructing an infinite volume equilibrium state}\label{sect:infvolequilibrium}

The mathematical construction of the equilibrium state $\tauglobal$ associated with an infinitely extended system  proceeds via two steps:  First, one takes the {\it ``thermodynamic limit''} for the bath $\b$, resulting in continuous spectrum of $\HB$, and a characterisation of its own equilibrium state $\tau_\b$ at inverse temperature $\beta$. Second, the much ``smaller'' system is coupled to the bath, which results in a new system plus bath equilibrium state $\tauglobal$.

We now explain the first step  for a bath consisting of free bosons with Hamiltonian $H_V$, see \eqref{eq:nparthamilt}. The procedure was first carried out in Ref.\cite{Araki1963} and further explained in Ref. \cite{Merkli2006, Joye2016} Consider a finite volume $V\subset{\mathbb R}^3$ of position space, say a cube of side length $L$, centered at the origin. The momenta $k_j$ and eigenstates $|\Psi_j\rangle$ of a single particle are explicitly known (the single particle Hamiltonian is just the Laplacian $-\Delta$; set $m_j=1/2$ in \eqref{eq:nparthamilt}). 

By applying a suitable selection of creation operators $a^\dagger(\Psi_j)$ [c.f. definition after \eqref{m18}] to the vacuum state $|0_V\rangle$ ($V$ indicates finite volume), one builds 
\begin{equation}
\label{42}
|\Psi_V\rangle = \frac{1}{\sqrt{n_1!\cdots n_p!}} \, a^\dagger(\Psi_1)^{n_1}\cdots a^\dagger(\Psi_p)^{n_p}\, |0_V\rangle,
\end{equation}
which is the state describing $n_1$ particles of momentum $k_1$ and $n_2$ particles of momentum $k_2$, and so on, in the volume $V$. Now one increases the volume $V$, keeping the density $\mu_j=n_j/|V|$ fixed, in such a way that the discrete distribution $\mu_j$ tends to a pre-selected function $\mu(k)$ of continuous momenta $k\in\mathbb R$. This $\mu(k)$ is called the (continuous) momentum density distribution because $\mu(k) \d k$ is the number of particles per unit volume (in position space) having momenta in the volume $\d k\subset {\mathbb R}^3$ around $k$.

As it turns out, the limit cannot be taken directly on the states. Rather, it has to be taken on averages of local observables $A$, which are operators built from (e.g. integrals of) creation and annihilation operators $a^\dagger_x$, $a_x$ with $x\in V$ for some finite (but arbitrary) $V\subset {\mathbb R}^3$ (the $a^\dagger_x, a_x$ are the Fourier transforms of $a^\dagger_k, a_k$). 
This limiting procedure defines the average $\langle A\rangle_{\beta,\infty}$ of the observable $A$ in the infinite volume state. The values of the expectation functional $A\mapsto \langle A\rangle_{\beta,\infty}$, for all observables $A$, define the infinite volume state. Now the question is how to represent this state as a vector (or density matrix)  $\tau_\b$. One can find a suitable Hilbert space ${\mathcal H}$ and a normalized state $|\Omega \rangle \in{\mathcal H}$, such that $\langle A\rangle_{\beta,\infty} = \langle \Omega| \pi(A) \, \Omega\rangle_{\mathcal H} = {\rm tr}_{{\mathcal H}}[|\Omega\rangle\langle\Omega|\pi(A)]$ (inner product and trace of $\mathcal H$). Here, $\pi$ is a {\it representation} of the observables, mapping each $A$ to an operator $\pi(A)$ acting on $\mathcal H$. The vector $|\Omega\rangle$, or equivalently, the density matrix $|\Omega\rangle\langle\Omega|$, is often times called the {\em purification} of the infinite volume state.  The triple $({\mathcal H}, \pi,\Omega)$ is called the {\it Gelfand-Naimark-Segal} representation \cite{BratteliRobinson1981, Haag1996}. It is given explicitly as follows, for any prescribed momentum density distribution $\mu(k)$.\cite{Araki1963, Merkli2006, Merkli2020} The Hilbert space is ${\mathcal H}={\mathcal F}\otimes{\mathcal F}$, where $\mathcal F$ is the usual Fock space for the bosonic gas, in which a general $N$ particle state is given by 
\begin{equation}
\label{43}
|\Phi_{\mathcal F}\rangle = \int_{{\mathbb R}^{3N}} \d k_1\cdots \d k_N \, \Phi(k_1,\ldots,k_N)  \, a^\dagger_{k_1}\cdots a^\dagger_{k_N}\, |0_{\mathcal F}\rangle,
\end{equation}
where $\Phi(k_1,\ldots, k_N)$ is the $N$-particle wave function in momentum representation. (Note that in \eqref{43} we integrate over all the possible continuous values $k_1,\ldots,k_N\in\mathbb R^3$ of the momenta of the $N$ particles; $k_1,\ldots,k_N$ are just integration variables, not to be confused with the fixed momenta chosen to build \eqref{42} in the finite volume situation.) The infinite volume bath state associated to the momentum density distribution $\mu(k)$ is represented as the vector $|\Omega \rangle :=| 0_{\mathcal F}\rangle \otimes |0_{\mathcal F} \rangle$, 
where $|0_{\mathcal F}\rangle$ is the vacuum state of the Fock space $\mathcal F$. The representation map is given by
\begin{equation}
    \pi(a^\dag_k) 
    = \sqrt{1+\mu(k)} \,\, a^\dag_k \otimes {\mathbf 1}_{{\mathcal F}} +  \sqrt{\mu(k)} \, \, {\mathbf 1}_{{\mathcal F}} \otimes  a_k.   
\end{equation} 
The desired density $\mu (k)$ is correctly reproduced, as one can check easily: $\langle \Omega| \pi(a^\dag_k a_{\ell}) \, \Omega\rangle_{\mathcal H} = \mu(k) \delta(k-\ell)$. The construction works for all momentum density distributions $\mu(k)$. Upon choosing Planck's black body distribution, $\mu(k) = 1/(e^{\beta \omega(k)}-1)$,
the state $|\Omega \rangle$ is the (purification of the) infinite volume  equilibrium state $\tau_\b$ for the bosonic gas in ${\mathbb R}^3$.

\smallskip

We now discuss the second step -- introducing the (much smaller) system. For an uncoupled $\s\b$ complex, with the infinitely extended $\b$, the global equilibrium state is $\rho\otimes |\Omega\rangle\langle\Omega|$, where $\rho\propto e^{-\beta \HS}$. The full (interacting) $\s\b$ equilibrium state, corresponding to an $\s\b$ interaction operator $V_{\s\b}$, see \eqref{eq:Htot}, is given by  \cite{BratteliRobinson1981, Derezinski2003, Merkli2020}  
$\tau_{\s\b} \propto e^{-\beta(L_0+\lambda\pi(V_{\s\b}))/2} (\rho\otimes|\Omega\rangle\langle\Omega|) e^{-\beta(L_0+\lambda\pi(V_{\s\b}))/2}$. Here, $L_0=L_\s+L_\b$ is called the non-interacting Liouville operator, with $L_\s\rho=[H_\s,\rho]$ (defined on system density matrices $\rho$) and $L_\b=H_\b\otimes{\mathbf 1}_{\mathcal F} - {\mathbf 1}_{\mathcal F}\otimes H_\b$ (acting on the purification Hilbert space ${\mathcal F}\otimes{\mathcal F}$), where $H_\b=\int_{{\mathbb R}^3} dk \omega(k)a^\dagger_k a_k$. 

The construction of the evolution of the infinitely extended bath and system complex follows the same procedure as the above infinite volume limit. Now one takes the thermodynamic limit of the finite volume Heisenberg picture {\it evolution} of observables. The evolution is represented in the purified Hilbert space by $e^{-i t L}$ where $L$ is the Liouville operator. It plays the role of the Hamiltonian, but now this evolution acts in the infinite volume Hilbert space $\mathcal H$, whose vectors represent states. 

To conclude this somewhat technical section, we summarize: It is possible to explicitly construct the equilibrium state of a system-bath complex, for a bath that is infinitely spatially extended. The state is represented by a vector in a new Hilbert space, not simply Fock space $\mathcal F$. In a way, when taking the volume of the bath to infinity and keeping the density of particles fixed and not zero, the usual Fock space is {\em not} suitable any longer to describe the equilibrium state. This is so because any density matrix acting on Fock state describes a state with only finitely many particles, which means a zero density at infinite volume! The explicit form of the infinitely extended $\s\b$ equilibrium state is an important ingredient in the rigorous analysis of the dynamics, such as for return to equilibrium, see Sec.~\ref{sect:RtE}.
In the following we will still use the notation $\tauglobal$  given in \eqref{eq:tausb}, even if we mean that the limiting procedure has been performed.
We further point out that this construction works for any value of the coupling parameter $\lambda$; no weak coupling regime is needed here.

\subsection{Long time $\s\b$ asymptotics and RtE } \label{sect:RtE}

We say that the property of RtE holds for a class of $\s\b$ initial states $\mathcal S$ and a class of $\s\b$ observables $\mathcal O$ if
\begin{equation} 
    \lim_{t\rightarrow\infty} \, \, \tr \big[\rho_{\s\b}(t) \, \, A\big] 
    = \tr \big[\tauglobal \, A\big],
\label{m3}
\end{equation}
for all initial states $\rho_{\s\b}(0)\in \mathcal S$ and all observables $A\in\mathcal O$. 
In dynamical systems parlance, \eqref{m3} means that the Gibbs state $\tau_{\s\b}$ is dynamically attractive and has a basin of attraction containing the class of states $\mathcal S$. The convergence is measured by limits of expectations of $\s\b$ observables $A\in \mathcal O$. We cannot expect \eqref{m3} to hold for all states and all observables. For instance, the initial state $\tau'_{\s\b}$, the equilibrium at a different temperature ${\beta'\neq \beta}$, is stationary and so it does not converge to $\tau_{\s\b}$ (nor does any other initial state approaching $\tau'_{\s\b}$ in the long time limit). Also, for models in which the bath is a spatially extended physical system, bath observables which sample space locations arbitrarily far away will capture deviations from the equilibrium state at arbitrarily late moments in time and so $\mathcal O$ should exclude such global observables. 

In the pioneering papers \cite{Jaksic1996, Bach2000,   Merkli2007-1,  Merkli2008, Merkli2008-1, Merkli2001, Frohlich2004} the property of RtE is shown to hold for an arbitrary $N$-level system coupled to a spatially infinitely extended bath of non-interacting bosons (as explained in Sec.~\ref{sect:infvolequilibrium}).
The $\s\b$ coupling is $\lambda V_{\s\b}$ (see \eqref{eq:Htot}), with $V_{\s\b}$ as in \eqref{m18}. 
The class of initial states $\mathcal S$ consists of all states that can be obtained by a local modification of $\tau_{\s\b}$  and the observable algebra $\mathcal O$ contains all system observables and all spatially localized bath observables. It is shown that \eqref{m3} holds provided the coupling constant $\lambda$ in \eqref{m1} is small enough, cf. \eqref{eq:binomialApproxConditionmain}, namely $0<|\lambda|<\lambda_0$ for some (not very explicit) $\lambda_0$. As the temperature $T$ becomes smaller, the upper bound $\lambda_0$ on $\lambda$ shrinks, and the method breaks down in the zero temperature case.  Beyond the smallness condition on $\lambda$, there are two further assumptions: the bath correlation function decays in time (exponential decay is assumed in the above references while in subsequent improvements \cite{Merkli2021a, Merkli2021b, Merkli2021c} polynomial decay suffices), and the so-called {\it Fermi Golden Rule Condition} is assumed ($\s$ and $\b$ are well coupled so that relaxation effects are visible at $O(\lambda^2)$).
Two remarks are in order: (i) The result \eqref{m3} holds for small enough $\lambda$, but the final state is {\em exactly} the coupled equilibrium state $\tau_{\s\b}$, to {\em all orders in $\lambda$}. (ii) The result \eqref{m3} is a statement about the full $\s\b$ dynamics, not merely the reduced system dynamics.

\bigskip

The class of $\s\b$ observables $\mathcal O$ contains ${\mathcal O}_\s$, the algebra of all observables acting on the system $\s$ alone. For the system dynamics, the primary consequence of the RtE relation \eqref{m3} is that for all system observables $A_\s$, 
\begin{equation}\label{m8}
    \lim_{t\rightarrow\infty} \tr \big[\rho_{\s\b}(t) \, A_\s \big] 
    = \trS[\tmf \, A_\s],
\end{equation}
with $\tmf$ the MFG state defined in  \eqref{eq:MFGibbs} for the total Hamiltonian \eqref{eq:Htot}. 
A special class of initial states for which \eqref{m8} holds is that of uncorrelated states of the form
\begin{equation}\label{m10}
    \rho_{\s\b}(0)=\rho\otimes\tau_{\b},
\end{equation}
where $\rho$ and $\tau_{\b}$ are, respectively, an arbitrary system density matrix and the bath equilibrium state. 
In other words, under the conditions mentioned above, namely that $0<|\lambda|<\lambda_0$, that the bath correlation function decays and that the Fermi Golden Rule holds, Refs.\cite{Jaksic1996, Bach2000, Merkli2001, Frohlich2004,Merkli2021a, Merkli2021b, Merkli2021c} show the following: For any global initial state \eqref{m10}, regardless of the details of $\rho$, the system  converges in the long time limit to the mean force Gibbs state $\tau_{\rm MF}$. The same holds for correlated initial states $\rho_{\s\b}(0)$ which are not of the product form \eqref{m10}, as long as they stay in the class $\mathcal S$ of states explained at the beginning of Sec.~\ref{sect:RtE}.

\subsection{Relation to non-integrable baths and eigenstate thermalization hypothesis (ETH)}
\label{sect:ETH}

In Subsection~\ref{Sec:OnFreeBath}, we outlined how the  bosonic bath can serve as a phenomenological model that can describe more complex, potentially non-integrable, baths. This raises the issue of comparing the system-bath approaches with those actively studied in many-body physics \cite{Gogolin2016,DAlessio2016,Mori2018,Deutsch2018}. 
The question of convergence to some equilibrium state (and in particular, thermalization to a thermal state) and the identification of the correct form of this state are  central questions also in this field.

Many-body physics usually considers systems of many identical particles.
Typically, each particle interacts with its neighbouring particles, either in  physical space (for example, particles in a gas) or in a lattice (for example, spin chains). 
Such many-body models are often non-integrable \cite{Caux2011,Lychkovskiy2020}. In this context, the celebrated ETH \cite{vonNeumann1929,Deutsch1991,Srednicki1994,Srednicki1999} answers the question about the steady state of a small subsystem as a part of a large isolated system. 
The ETH can be viewed as a quantum version of the ergodic hypothesis in classical mechanics. Though there is no rigorous proof of the ETH (the same holds for the ergodic hypothesis for most classical models), there is enormous numerical evidence that the ETH is satisfied for a large class of non-integrable physical models.

A bath consisting of non-interacting particles is integrable and does not obey the ETH.
It is not known whether the 
complex, obtained by coupling this bath to a system, satisfies the ETH.
There are studies which show that a localized perturbation of an integrable system is often sufficient to obtain a non-integrable and thermalizing system \cite{Znidaric2020,Brenes2020,LeBlond2021,Schonle2021}, while other results shows that this is not always the case \cite{Fagotti2017}.
Note that the non-interacting bosonic bath coupled to a system, as detailed in section \ref{subsec:discbosonic}, can be exactly mapped into a chain of interacting harmonic oscillators coupled to the system \cite{Chin2010a,Prior2010}. This observation could serve as a starting point to compare and link the bosonic bath used in open systems theory with many-body physics baths.

If we use the model of non-interacting bosons as a phenomenological model for a more complex non-integrable bath satisfying ETH, then one may argue that our question (Q) about the steady state is answered directly by the ETH. 
However, just like the ergodic hypothesis in classical mechanics, the ETH does not say anything about the {\it rate} of equilibration. In contrast, with the bosonic bath, one does obtain this more detailed information, including decoherence rates. One may then ask whether the non-integrability of the bath is responsible for thermalization, or whether thermalization can be explained  from the viewpoint of a {non-interacting} bath model? Imagine a very weakly non-integrable bath, for which one may expect that thermalization process caused by the ETH occurs very slowly. However, it might be that thermalization occurs much faster due to a mechanism independent of non-integrability and the ETH, a mechanism which can be explained and understood in the framework of a simplified model of a non-interacting bosonic bath. It would be interesting to compare the predictions of both models and to establish further links between them in future works.

\section{System dynamics and steady state} \label{sec:dynamicsB}

The results presented in Sec.~\ref{sect:RtE} are concerned with the asymptotics of the full $\s\b$ dynamics at $t\rightarrow\infty$ leading to \eqref{m3}, which also implies that the system converges to the stationary state $\tau_{\rm MF}$. Now we consider the microscopic details of the dynamics of the system alone, starting from the  combined system and bath complex evolving according to the unitary dynamics given by the total Hamiltonian $\Htot =\HS + \HB + \lambda \, V_{\s\b}$, as stated in Eq.~\eqref{eq:Htot}.

\subsection{General dynamical setting} 
\label{sect:Sevol}

The dynamical point-of-view for an {\it open system $\s$} is generally concerned with describing its state evolution $\rhoSt$ when $\s$ is brought into contact with a heat bath $\b$ at temperature $T$. One tries to solve the dynamical equations of motion for the system, or at least to determine the system's steady state  $\stst = \lim_{t \to \infty}\rho(t)$, cf. \eqref{convtoss}. One may then address  a more refined version than our initial question (Q): 
\begin{itemize} 
\item[(Q')] Is the system's dynamical steady state $\stst$ equal to the mean force Gibbs state $\tmf$ discussed in the statics section?
\end{itemize}
Below we summarize some key results on steady states of various dynamical systems, and make a connection to $\tmf$ where possible.

To proceed, consider the general interaction of the form \eqref{vint}. If the $X^{(j)}$ commute with $\HS$, then the interaction is called {\it energy conserving}. The system populations are constant in time but the bath still causes irreversible effects in the system, such as decoherence. Those models are suitable for situations in which decoherence happens much more quickly than thermalization and we are interested only in the decoherence time scale.\cite{Palma1996, Merkli2018} If at least one $X^{(j)}$ does not commute with $\HS$ then energy exchange processes between the system and bath are enabled.

The initial $\s\b$ state, $\rho_{\s\b}(0)$, is usually assumed to be of product form \eqref{m10}, although the evolution of correlated or entangled initial states is also relevant and studied \cite{Karrlein1997,Buser2017,Alipour2019,Paz-Silva2019, Merkli2021c,Gorini1989,Trevisan2021,Trush2021}.
The state of $\s$ at time $t$ is the reduced density matrix, cf. \eqref{m2}, 
\begin{equation}\label{eq:systemstate}
	\rhoSt = \trB[\rho_{\s\b}(t)]
	= \trB[ e^{-i t \Htot} \, \rho_{\s\b}(0)\,  e^{i t \Htot}].
\end{equation}
Taking the time-derivative gives
\begin{equation}\label{eq:GeneralME}
	\dotrhoSt  = -i \, \, \trB \big[ \,  [H_{\text{tot}},\rho_{\s\b}(t)] \, \big],
\end{equation}
from which we want to derive an autonomous equation, a {\it master equation}, for $\rhoSt$. 
For product initial states \eqref{m10}, equation \eqref{eq:GeneralME} can be cast into the form of an integro-differential equation for $\rhoSt$, which is called the {\it Nakajima-Zwanzig master equation} \cite{BrPet2002}.
For an arbitrary initial state $\rho_{\s\b}(0)$, where $\s$ and $\b$ are correlated or entangled,  the master equation for $\rhoSt$ will depend on the initial $\s\b$ correlation, see e.g. Refs.~\cite{Karrlein1997,Romero1997,Alipour2019, Mori2008, Tasaki2007, Yuasa2007, Vacchini2016a, Merkli2021c,Gorini1989,Trevisan2021}.

With very few exceptions, a master equation for $\rhoSt$ (in an analytically closed form)  cannot be derived without recourse to approximations. In the sections below we will discuss various approximations in the well-studied weak coupling limit, as well as the ultrastrong and intermediate limit. Before we do so, we will introduce some useful terms and notations, and then begin the discussion with the exactly solvable model of the open dynamics of the quantum harmonic oscillator.

\bigskip

Unless otherwise stated, the analysis of the master equations discussed in the following sections assumes the Hamiltonian $H_\text{tot}$ defined in \eqref{eq:Htot} with a single interaction term $X \otimes B$ in \eqref{vint}.  We will make use of decomposition (23) of the operator $X$ into a sum of energy eigenoperators $X_m$.
In the interaction picture (denoted by the tilde), one has  $\tilde{X}_m(t)=X_m \, e^{i\omega_mt}$. These quantities define two timescales, the one associated to the Bohr frequencies $\omega_m$, the other one associated to the differences of Bohr freuencies, $\omega_{mn}=\omega_{m}-\omega_n$.

Furthermore, the {\it bath correlation function} $G(t)$ plays a central role in open system dynamics, particularly at weak coupling, as it determines the memory time of the $\s\b$ interaction. It is defined as the auto-correlation function of the bath operator $B$, 
\begin{equation}\label{eq:ResCorrFunc}
	G(t)= \trB\left[e^{-i t \HB} \, B \, e^{i t \HB } \, B \, \tau_B\right],
\end{equation}
and is the well-known dynamic version of the imaginary-time bath correlation function introduced in Sec.~\ref{subsec:staticpolaron}. It satisfies the Kubo-Martin-Schwinger (KMS) condition $G(t)=G(-t-i \beta)$ where $\beta$ is the inverse temperature of the bath Gibbs state $\tau_\b$. $G(t)$ is assumed to vanish as $t\to\infty$ and this decay defines the time scale, $t_\b$, of the correlations of the quantum noise. 
We also define the time dependent coefficients 
\begin{equation} \label{eq:Gammam}
    \Gamma_m(t)=\int_{0}^{t} \d r \, e^{-i\omega_m \, r} \, G(r) \quad \mbox{with} \quad \Gamma_m\equiv\Gamma_m(\infty),
\end{equation}
which will appear in the master equations below.

\subsection{Exact dynamics of the damped quantum harmonic oscillator} \label{subsec:dynqho}

A general method to solving \cite{Grabert1988} the exact dynamics of a damped harmonic oscillator described by \eqref{eq:CLHam} follows a path integral functional integral approach. For general spectral density $J(\omega)$ and coupling strength $\lambda$, and a broad class of initial states of the global system, including the global Gibbs state $\tauglobal$ and entangled states, it is shown \cite{Grabert1988} that their steady states are all the same. The initial state $\tauglobal$ is clearly a stationary state of the global evolution, and its reduced system state is $\tmf$. Hence $\tmf$ {\it is} the steady state for the whole class of initial states considered. This state is given in Eq.~\eqref{eq:harmonicoscitmf} in the statics section \ref{sec:statics}.

An alternative method to show the dynamical convergence to $\tmf$ is obtained in Ref.~\cite{Fleming2011b}  using Heisenberg-Langevin equation methods. 
More recent work \cite{Subasi2012} addresses general $N$-body quantum Brownian motion. 
Using  Heisenberg-Langevin equation methods, the general two-time correlation functions are explicitly evaluated \cite{Subasi2012}.  This is done for initial states $\rho_{\s\b}(0)=\rho(0)\otimes \tau_\b$ c.f. \eqref{m10}, and in the steady state limit of long times $t\to\infty$.
Then the above stationary state argument \cite{Grabert1988}  for $\tauglobal$ is again applied, to prove  that the steady state of the $N$-body system, for initial states $\rho_{\s\b}(0)=\rho(0)\otimes \tau_\b$,  must coincide with the MFG state $\tmf$.


Working from the functional integral description of the system dynamics, it is also possible to construct an exact non-Markovian master equation \cite{Hu1992}, and specify its steady state in terms of its Wigner function \cite{Ford2001}. A generalization to the case of the driven damped quantum oscillator has been also obtained \cite{Qiu2021}.

\smallskip

Taken together the studies above firmly establish, that the dynamical steady state $\stst$ of a quantum oscillator, under dynamics given by \eqref{eq:CLHam} for a broad class of initial global states, is \textit{exactly} the mean force Gibbs state $\tmf$, at all system-bath coupling strengths $\lambda$ and for general spectral density $J(\omega)$.

\subsection{Weak coupling dynamics} \label{subsec:weakdyn}

In most cases, approximate forms for the master equation are obtained on the assumption that the system-bath coupling is weak enough that a second order perturbative treatment suffices. This defines the `weak coupling limit' for master equations, as indicated in Fig.~\ref{fig:weakvsstrong}. Weak coupling applies  in a variety of contexts, including  quantum optical systems \cite{CohenTann1992,Carmichael1993,Carmichael1999,Agarwal2012,Cottet2017}, nuclear magnetic resonance\cite{Wangsness1953,Redfield1957,Redfield1965,Haeberlen1976,Abragam1983,Mehring1983,Ernst1990,Kowalewski2017,Bengs2020,Bengs2021}, solid state and molecular physics \cite{MayKuhn,Valkunas2013,Ramsay2010,Ramsay2010a}, and in certain biological systems\cite{Rebentrost2009a,Olaya2008,Fassioli2010,Abramavicius2011,Hoyer2014,Dodin2018,Novoder2017}. Weak coupling expansions generally lead to master equations that are relatively simple and have useful immediate physical interpretations.
For example, they predict  basic properties of open quantum systems such as decoherence and thermalization\cite{Alicki2007,BrPet2002,AccardiBook,AccardiKozyrev,Trush2019,Fagnola2019,Lostaglio2015}, as well as more complicated properties such as laws of thermodynamics\cite{McAdory1977,Spohn1978,Spohn1978a,Alicki2018,Kosloff2013,Potts2019}, environment-assisted quantum transport\cite{Rebentrost2009,Mohseni2008,Caruso2009,Chin2010,Chin2012}, superradiance and supertransfer\cite{Abasto2012,AVK2015}, emergence of dark states\cite{Zhang2016,Volovich2016,Hu2018}, decoherence-free subspaces\cite{Agredo2014}, etc. 
Even if a weak-coupling approximation is not strictly valid,
many important system properties 
can be  captured  qualitatively in this approximation. 
Often, phenomenological equations of the GKSL form\cite{GKSLform} are used\cite{Olaya2008,Cottet2017}, which implicitly assume weak coupling. 
Moreover, for the example of light-harvesting complexes, the system (representing the electronic DoFs) is often strongly coupled only to a finite number of distinguished vibrational modes\cite{Kolli2012}. 
If these are included into an enlarged system, then the weak coupling approximation can be applied to the enlarged system now interacting with the remainder of the bath\cite{Plenio2013,Novoder2017}.

The weak coupling limit allows for the introduction of two important approximations. The first is the \textit{Born approximation} which is based on the assumption that, as far as the system dynamics is concerned, any changes in the state of the bath, or any correlations that might develop between system and bath due to their interaction\cite{BrPet2002,Tasaki2007, Yuasa2007, Merkli2021c}, can be neglected. 
The second approximation is the {\it Markov approximation}, in which the rate of change of the state of the system at time $t$ is assumed to depend only on its  state at that same time, and not on its history. The Born  approximation is justified due to the difference of the sizes of $\s$ and $\b$ (small influence of $\s$ on $\b$), the Markov approximation is justified via a separation-of-timescale argument \cite{BrPet2002}. 
These approximations are almost universally inserted into derivations of master equations, for example, for the Bloch-Redfield ME discussed in Sec.~\ref{Sec:BRME}.

\subsubsection{\wk~Davies theory and resonance theory}
\label{sec:DaviesArgument1}
A rigorous analysis of the weak coupling limit has been provided in pioneering work by Davies \cite{Davies1976, Davies1974, Davies1976a}. The master equation is derived in the {\it Bogoliubov-van Hove limit} \cite{Bogol1945,vanHove1955}, where  $\theta=\lambda^2t$ is taken as constant, but  $\lambda\to0$ and  simultaneously $t\to\infty$. The rescaled time $\theta$ is sometimes called the coarse-grained time.
For initial product states $\rho \otimes \tau_\b$, and under certain conditions on the bath correlation function $G(t)$, a master equation can then be derived rigorously, without making the explicit Born-Markov approximation.
Specifically, Davies shows that the reduced system dynamics $\tilde \rho(\theta)$ (interaction picture indicated by $\tilde{\phantom{\cdot}}$\,) is well approximated by some $\tilde \rho^\Dav(\theta)$, i.e. 
\begin{equation}\label{m13}
    \lim_{\lambda\to 0}\ 
    \| \tilde\rho(\theta) - \tilde\rho^\Dav(\theta)\| =0,
\end{equation}
where the convergence is uniform on the segment $\theta\in[0,\Theta]$ for an arbitrary finite $\Theta$ and  $\| \cdot \|$ denotes the trace norm.
Here $\tilde\rho^\Dav(\theta)$ is the solution to the {\it Davies master equation} in  interaction picture and for the coarse-grained time $\theta = \lambda^ 2 t$,    \cite{Davies1976, Davies1974, Davies1976a}
\begin{equation}\label{eq:DaviesME}
	\frac{d\tilde{\rho} (\theta)}{d\theta} 
	= \LD  \, \tilde \rho (\theta),
\end{equation} 
where $\LD$ does not depend on $\theta$, and is given by\footnote{The Davies master equation in \eqref{eq:DaviesME} is of GKSL form \cite{GKSLform}.}
\begin{equation} \label{Daviesgenerator}
   \LD \,  \tilde{\rho}
    = 	-i\left[\Delta H^{||}, \tilde{\rho} \right] 
    +\sum_{m}\gamma_m \, \left(X_m  \, \tilde{\rho} \,  X_m^\dag-\tfrac{1}{2}\left\{X_m^\dag  X_m, \, \tilde{\rho} \right\}\right).
\end{equation}
The positive damping rates are given  by $\gamma_m=2 \, \text{Re}[\Gamma_m]$, and the term that renormalises the system energy is 
\begin{equation} \label{eq:DaviesHren}
	\Delta H^{||} = \sum_{m} \, \text{Im}[\Gamma_m] \, X_m^\dag X_m,
\end{equation}
where $X_m$ and $\Gamma_m$ are the energy eigenoperators and the master equation coefficients defined in \eqref{m16} and  \eqref{eq:Gammam}, respectively. 
Note that this renormalisation commutes with the system's bare Hamiltonian $\HS$, i.e. $\left[H_\s,\Delta H^{||}\right]=0$.

\smallskip

We now discuss the implications of the convergence \eqref{m13} at coarse-grained times $\theta \ge 0$. If one assumes non-degenerate energy levels of $\HS$ and non-degenerate non-zero frequency differences  $\omega_{mn}$, see \ref{sect:Sevol},  then equation \eqref{Daviesgenerator} predicts no coupling between system populations (diagonal entries of $\tilde \rho$ in the $H_\s$ eigenbasis) and the coherences (off-diagonals). 
The coherences are found to decay to zero as $t\to \infty$. 

The above assumptions imply that $\gamma_m$ satisfy the detailed balance conditions $\gamma_m=e^{-\beta\omega_m}\gamma_{-m}$. As a consequence, the steady state of equation \eqref{eq:DaviesME} is then readily shown \cite{Davies1976, Davies1974, Davies1976a} to be the system  Gibbs state $\tau$, cf. Eq.~\eqref{eq:SGibbs}. 
For degenerate energy levels the steady state is non-unique \cite{AVK2015,Lostaglio2015,Garcia2018,Trush2017,Trush2019,Merkli2015} and the situation becomes more complex.   

\smallskip

Thus to summarise, working in the Bogoliubov-van Hove limit and assuming the generic non-degenerate case, Davies showed that the steady state is the standard Gibbs state $\tau$, confirming the validity of Gibbs statistical physics, see Fig.~\ref{fig:weakvsstrong}. 
However, this jars with the static assumption of the steady state being the MFG state $\tmf$, which  includes ${\mathcal O}(\lambda^2)$ and higher order corrections in comparison to $\tau$, see \eqref{eq:tauMFseries}. There are a number of reasons for this incongruence.

\smallskip

Firstly, in order to obtain an equation in so-called {\it GKSL form} \cite{GKSLform,GKS1976,Lindblad1976,Alicki2007,Franke1976,Chruscinski2017}  in his derivation of \eqref{eq:DaviesME},  Davies makes the so-called {\it secular approximation}, (see Sec.~\ref{subsubsec:otherweak}), which can remove dynamical features leading to the final state $\tmf$, and forcing instead convergence to $\tau$. 
Secondly, the Davies master equation is of second order in $\lambda$, but relaxation also takes place on timescales of order $\lambda^{-2}$. Small errors accumulate over long times and thus the steady state is trustworthy only up to the zeroth order.
Thirdly, the convergence in Eq.~\eqref{eq:DaviesME} is proved for arbitrary large but finite rescaled time intervals $\theta\in[0,\Theta]$. Davies' proof  requires $\lambda$ to be smaller and smaller as $\Theta$  increases. Hence, strictly speaking, it is not guaranteed that the steady state of the Davies master equation coincides with the true steady state in the weak-coupling limit. 

\bigskip

These limitations are overcome by the {\em quantum resonance theory}, which improves Davies' results. Namely, under the same conditions used to show RtE (Sec.~\ref{sect:RtE}), it is shown in Refs.\cite{Konenberg2017, Merkli2020,Merkli2021a, Merkli2021b, Merkli2021c} that \eqref{m13} can be improved to
\begin{equation}\label{m21.1}
\| \tilde\rho(t) - \tilde\rho^\Dav(t)\|  \le C \, \lambda^2,
\end{equation}
valid for all times $t\ge 0$ (no coarse graining necessary), where $C$ is a constant independent of $\lambda$ and $t$. Equation \eqref{m21.1} shows that the solution of the Davies master equation approximates the system dynamics to ${\mathcal O}(\lambda^2)$ {\em at all time scales}. In particular, this guarantees that the system Gibbs state $\tau$, which is the steady state predicted by the Davies master equation, is the true steady state up to ${\mathcal O}(\lambda^2)$. This is entirely consistent with the result of RtE, saying that the stationary state is $\tau_{\rm MF}=\tau +{\mathcal O}(\lambda^2)$. It is further shown in \cite{Konenberg2017, Merkli2020} that by adding to the Davies generator $\tilde{\mathcal L}^{\rm D}$ higher order terms
\footnote{The higher-order terms and the whole generator $\tilde{\mathcal L}^{\rm KM}$ are also of GKSL form \cite{GKSLform}.}, 
\begin{equation}\label{eq:KM}
\tilde{\mathcal L}^{\rm KM} =\tilde{\mathcal L}^{\rm D} + \lambda \tilde{\mathcal L}_1+\lambda^2 \tilde{\mathcal L}_2+\cdots,
\end{equation}
the solution $\tilde\rho^{\rm KM}(t)$ of the corresponding master equation $\tfrac{d}{dt}\tilde \rho^{\rm KM}(t) = \tilde{\mathcal L}^{\rm KM} \rho^{\rm KM}(t)$ has the following properties. Firstly, it is asymptotically exact, meaning that the stationary state is the exact mean force Gibbs state, $\lim_{t\rightarrow\infty} \tilde\rho^{\rm KM}(t) =\tau_{\rm MF}$. Secondly, the populations (diagonal density matrix elements in the $H_\s$ basis) of the system are approximated by those of $\tilde\rho^{\rm KM}(t)$ to ${\mathcal O}(\lambda)$ for all times $t\ge 0$. The coherences of the system, (off-diagonals) are  guaranteed to be approximated by those of $\tilde\rho^{\rm KM}(t)$ to ${\mathcal O}(\lambda)$ for times $t$ in the windows $\lambda^2t <C_1$ and $\lambda^2 t>C_2$ for some constants $C_1$, $C_2$.  The corrections $\tilde{\mathcal L}_k$ are constructed by an explicit perturbation procedure. 

We close this section by mentioning that in the literature, perturbation theory is usually carried out by simply neglecting higher order terms, without controlling their size for large times. In contrast, the resonance theory is a rigorous approach, in which remainder terms are estimated to be small, for all times.

\subsubsection{\wk~Bloch-Redfield master equation (BRME)} \label{Sec:BRME}

For $\Htot$ with a single interaction term $\lambda \, X \otimes B$, the general form of the second order ($O(\lambda^2)$)  master equation obtained by invoking the Born and Markov approximations is known as the {\it Bloch-Redfield master equation} \cite{Wangsness1953,Bloch1957,Redfield1957,Redfield1965,BrPet2002}, 
\begin{eqnarray} \label{eq:BRMEGeneratorDefined}
	\dt{\rho} = \mathcal{L}_t^\text{BR} \left(\rho\right)
	&=& -i\left[H_\s+\lambda^2 \Delta H^{||}(t)+\lambda^2 \Delta H^\perp(t), \rho \right]\\
	&&+\lambda^2 \sum_{mn}\gamma_{mn}(t)\left(X_m\rho X^\dagger_n-\tfrac{1}{2}\left\{X_n^\dagger X_m,\rho \right\} \right). \nonumber
\end{eqnarray}
The energy renormalization contributions consist of a contribution that commutes with $\HS$,
\begin{equation} \label{eq:parallelH}
	\Delta H^{||}(t)= \sum_{m}\text{Im}\left[\Gamma_m(t)\right] \, X_m^\dag X_m,
\end{equation}
and one that generally does not commute with $\HS$,
\begin{equation} \label{eq:orthongonalH} 
	\Delta H^\perp(t)=\tfrac{1}{2}\sum_{m\ne n}\left(\Gamma_m(t)-\Gamma_n^*(t)\right)X_n^\dagger X_m,
\end{equation}
where the $X_m$ and $\Gamma_m(t)$ are the  energy eigenoperators  and the master equation's damping coefficients defined in \eqref{m16} and  \eqref{eq:Gammam}, respectively. 

The second line in \eqref{eq:BRMEGeneratorDefined} contains the damping matrix 
\begin{equation} \label{eq:dampmatrix}
    \gamma_{mn}(t)=\Gamma_{m}(t)+\Gamma_n^*(t), 
\end{equation}
Note that, in general, \eqref{eq:parallelH}, \eqref{eq:orthongonalH}, and \eqref{eq:dampmatrix} are all time-dependent. 

The time dependence of the generator $\mathcal{L}_t^\text{BR}$ implies that the master equation is non-Markovian in the sense that the time evolution operator $\Lambda_t$, defined by $\rho(t)=\Lambda_t\rho(0)$, is such that $\Lambda_t\Lambda_s\ne\Lambda_{s+t}$, {\it i.e.}, it does not satisfy the semi-group property criterion for Markovianity \cite{Li2018}. This time dependence makes solving equation  \eqref{eq:BRMEGeneratorDefined} much more difficult than solving the Davies ME  \eqref{eq:DaviesME} for which $\LD$ is $\theta$-independent. 
However, due to the bath correlation decay,  the functions $\Gamma_m(t)$ saturate to constant values for times $t\gg t_\b$. In this regime, one can replace $\Gamma_m(t)$ by $\Gamma_m(\infty)\equiv \Gamma_m$ and obtain the time independent generator $\mathcal{L}_\infty^\text{BR}$. At large times, the generators $\mathcal{L}_t^\text{BR}$ and $\mathcal{L}_\infty^\text{BR}$ coincide, and since we are interested in steady-state solutions, we will always mean the  simpler, time-independent version $\mathcal{L}_\infty^\text{BR}$.

\bigskip

In the previous section we found that the steady state of the Davies master equation is the Gibbs state $\tau$, i.e. this steady state does not show any signature of the bath interaction other than the bath's inverse temperature $\beta$.
Having next introduced the Bloch-Redfield equation,  we are now in the position to discuss the complex world of its steady state(s).

\smallskip

A first setback is that, contrary to the Davies ME, there is no analytical expression for the steady state of the Bloch-Redfield equation. Two approaches concerning the issue of determining the steady state can be identified, which are discussed below. The first is the one adopted by Geva {\it et.al.} \cite{Geva2000} and in more detail by Mori and Miyashita \cite{Mori2008} in which the aim is to derive an expression for the time derivative $d\rho/dt$, and confirm to second order in $\lambda$ that the mean force Gibbs state is such that this derivative vanishes, i.e. $\tmf$ ``qualifies'' as the dynamical steady state. 
The second approach adopted, e.g., by Fleming {\it et al} \cite{Fleming2011}, Thingna {\it et al} \cite{Thingna2012}, Suba\c{s}{\i} {\it et al} \cite{Subasi2012} and Purkayastha {\it et al} \cite{Purkayastha2020} is to construct a master equation for $\rho(t)$, to search for its long time steady state solution to second order in $\lambda$, and to compare this with the mean force Gibbs state also evaluated to second order.

\smallskip

As an example for the first approach, Mori and Miyashita  \cite{Mori2008} consider the time derivative $\dotrhoSt$ to second order in $\lambda$ (weak coupling), which is given in terms of $\rho(t)$ and the total system state $\rho_{\s\b}(t_0)$ at an earlier time $t_0$. %
Like for the quantum oscillator case discussed in Sec.~\ref{subsec:dynqho}, this initial state is chosen as either (i)  $\rho(t_0)\otimes \tau_\b$ or (ii) the global  Gibbs state $\tauglobal$. 
Assuming now that $\rho(t)$ is set equal to the mean force Gibbs state $\tmf$  (also to second order in $\lambda$ see \eqref{eq:tauMFseries}), Mori and Miyashita confirm \cite{Mori2008} that the derivative $\dotrhoSt$ is zero (to second order in $\lambda$), for  both initial state choices (i) and (ii). I.e. 
$\mathcal{L}_\infty^\text{BR} \left(\tmf\right)=0 = \dt{\rho}$.
Interestingly, the same statement is recovered by letting $t_0\to-\infty$ while \textit{no constraint} is imposed on the form of the initial state in that infinite past. I.e. all memory of the initial conditions are lost. It is quite general then to view the above conclusion to hold {\it independently of the initial state of the total system}, at least to second order in $\lambda$. 

The above result \cite{Mori2008} is applicable for  a general quantum systems $S$ in contact with a bath at inverse temperature $\beta$, and is valid to second order in $\lambda$.  It confirms that $\tmf$ at inverse temperature $\beta$ is a steady state of the Bloch-Redfield dynamics $\mathcal{L}_\infty^\text{BR}$ given in  \eqref{eq:BRMEGeneratorDefined}.

Mori and Miyashita also note, that there is however an important ambiguity regarding the steady state as follows. 
The algebraic equation $\mathcal{L}_\infty^\text{BR}(\rho)=  \mathcal{O}(\lambda^3)$ for the steady state yields a steady state $\stst^{(2)}$ accurate to $\mathcal{O}(\lambda^2)$.
But for any second order generator $\mathcal{L}^{(2)}$, and for $W$ a traceless operator that is diagonal in the $\HS$ basis, one has $\mathcal{L}^{(0)}(W)= - i \, [\HS, W] =0$ and $\mathcal{L}^{(2)}(W)=\mathcal{O} (\lambda^2)$. 
Thus for states $\stst^{(2)}+\lambda^2 W$ one finds that they are also steady state solutions to the same order in $\lambda$ as $\stst^{(2)}$ itself, i.e. $\mathcal{L}_\infty^\text{BR}(\stst^{(2)}+\lambda^2 W) = \mathcal{O}(\lambda^3)$ due to the linearity of the evolution $\mathcal{L}_\infty^\text{BR}$ in the density matrix $\rho$, cf. Eq.~\eqref{eq:BRMEGeneratorDefined}.
This ambiguity implies that while the steady state $\stst^{(2)}$ is correct to second order in $\lambda^2$ for the {\em off-diagonal elements} in the $\HS$ basis, the {\em diagonal elements} are correct only to zeroth order. I.e. that is the order where it is correct to simply replace the diagonal elements of $\tmf$ with those of $\tau$, see \eqref{eq:tauMFseries}.

\smallskip

Fleming {\it et al} \cite{Fleming2011} (see also Ref.~\cite{Tupkary2021}) show that the above ambiguity in the steady state is a consequence of the perturbative approach used to construct the master equation: solving a $2n$ order perturbative master equation will yield the diagonal elements of the steady state accurate only to order $2n-2$ (while the off-diagonal elements are determined to order $2n$). So resolving the second order indeterminacy requires expanding the master equation to {\it fourth order}, and then using degenerate perturbation theory to find the needed corrections, thus fixing $W$ in the above. A slightly different proof of the last conclusion is given in Ref.~\cite{Tupkary2021}.

Thingna {\it et al} \cite{Thingna2012} reach a similar conclusion to Ref.~\cite{Fleming2011}, but bypass the need to calculate fourth order terms in the master equation, typically a very complex task. They do so by making use of analytic continuation methods to derive the corrections to the diagonal elements of the steady state based on its second order off-diagonal elements alone. They confirm that the resulting steady state solution, now correct to second order in $\lambda$ and uniquely defined, is identical to the mean force Gibbs state $\tmf$ to $\mathcal{O}(\lambda^2)$.  
Suba\c{s}{\i} {\it et al} \cite{Subasi2012}  generalise this result to the case where the system-bath interaction is of the more general form \eqref{vint}, while the initial state is restricted to the product state \eqref{m10}.
For a double-quantum-dot charge qubit, described by the spin-boson model \cite{Purkayastha2020}, the above results \cite{Mori2008, Fleming2011, Thingna2012, Subasi2012} on the convergence of the dynamics (to second order in $\lambda$) to the $\tmf$ (to second order in $\lambda$), are calculated and illustrated in detail in Ref.~\cite{Purkayastha2020}.

\smallskip 

The Bloch-Redfield master equation does have the problem of generating non-positive probabilities, at least for some early times. The trustworthiness of this master equation has been questioned  \cite{Rivas2011} for this reason, though inclusion of a slippage of initial conditions\cite{Suarez1992,Gaspard1999,Fruchtman2016} corrects this, without having an impact on the final steady state.
\footnote{As noticed in the mentioned papers and also in Ref.~\cite{Trush2021}, the violation of positivity at early times is caused by initial highly non-Markovian dynamics (see also Ref.~\cite{Tere2021}). This true dynamics, during which system and bath adjust to each other,  cannot be described by the BRME since it makes the Markovian assumption.}
But what we saw here is that its long time steady state, $\stst$, has corrections to $\tau$ that are partially consistent with  $\tmf$, in line with the expectation from the static point-of-view \ref{sec:statics}. Assuming the weak coupling limit is justified  (here meaning expansions in $\lambda^2$, as well as $\lambda^4$ to get the ${\mathcal O}(\lambda^2)$ corrections for the diagonals), it hence appears that the BRME is  trustworthy when it comes to predicting long time behaviour and equilibration. It also gives  well-behaved, positive predictions at shorter times when the initial system state is assumed to be close to the $\tmf$.

\smallskip

It is worthwhile to comment on the above results \cite{Mori2008,Subasi2012} in comparison to those of RtE, cf. Sec. \ref{sect:RtE}. 
Coincidence of the {\it perturbative expansion} of the steady state of the BRME with that of the mean force Gibbs state up to a finite order does not in fact guarantee that these states are exactly the same.  To contrast, the results on RtE prove the exact (i.e., in all orders of $\lambda$) equality  of these states. This is proven for any $\lambda$ smaller than a fixed strictly positive number $\lambda_0$. For this reason, the results of Refs.~\cite{Mori2008,Subasi2012} are most relevant in the context of discussions about the choice between the Bloch-Redfield and the Davies (secular Bloch-Redfield) master equation: Which one reproduces the true equilibrium state more precisely. If, for a concrete problem to solve, it is important to have a steady state that contains corrections to the  Gibbs state $\tau$, then the Bloch-Redfield equation or the master equation (\ref{eq:KM}) should be preferred.

\subsubsection{\wk~Further results} 
\label{subsubsec:otherweak}

A consequence of taking the long time form $\mathcal{L}_\infty^\text{BR}$ for the Bloch-Redfield generator is that the master equation is  
not of GKSL form\cite{GKS1976,Lindblad1976,Franke1976,Alicki2007, Chruscinski2017} as the damping matrix,  $\gamma_{mn}$, defined in Eq.~\eqref{eq:dampmatrix}, is not necessarily positive definite. Negative probabilities can now occur for some initial states of the system \cite{Gnutzmann1996,Alicki2007,Whitney2008}
at times less than $t_\b$, the bath correlation time. 
This weakness can be removed by making a further approximation, the secular approximation, already introduced by Davies, cf Sec.~\ref{sec:DaviesArgument1}, which involves removing those terms $X_m\rho X_n^\dagger$ in Eq.\ (\ref{eq:BRMEGeneratorDefined}) that oscillate, in the interaction picture, at frequencies $\omega_{mn}\gg t_\s^{-1}$ where $t_\s$ is the system relaxation time. The resultant full-secular-approximation form of the Bloch-Redfield master equation can be readily shown \cite{BrPet2002} to be of GKSL form, and moreover has the consequence that the steady state is the Gibbs state, $\tau$.
But if nearly degenerate Bohr frequencies are found to occur, then the corresponding non-secular terms in the master equation are important, and should not be removed.

It has been noted that the full secular approximation is quite often made indiscriminately, even in circumstances where it is not justified. This appears motivated, at least in part, by it leading to a GKSL master equation which has the Gibbs state as its steady state, thus connecting the ME dynamics to Gibbs state physics, see Fig.~\ref{fig:weakvsstrong}. If only those non-secular terms that satisfy $\omega_{mn}\gg t_\s^{-1}$ are removed \cite{Tscherbul2015,Jeske2015,Cresser2017,Cattaneo2019,Cattaneo2020,Farina2019a}, called partial secular approximation, the resultant master equation will still not be of GKSL form, and moreover the steady state is no longer guaranteed to be either the Gibbs state, or a second order approximation to the mean force Gibbs state.

A further approximation, often made in practice, is to ignore the imaginary parts of the master equation coefficients $\Gamma_m$, see Eq.~\eqref{eq:Gammam}. It turns out \cite{Mori2008,Agarwal2001} that the Bloch-Redfield equation with $\Gamma_m$ replaced by ${\rm Re}[\Gamma_m]$ has the Gibbs state $\tau$ as its  steady state.

Recently, a rigorous derivation of a {\em unified} GKSL quantum master equation beyond the secular approximation is proposed\cite{TrushUni}. A rigorous procedure leads, in general, to a partial secular approximation followed by certain modifications in the arguments of the spectral density function. Since the derivation is rigorous, the resulting equation is thermodynamically consistent, see also \cite{Latune2020,Potts2021}. In particular, a Gibbs state with respect to a modified system Hamiltonian (in which all nearly degenerate energy levels and Bohr frequencies become exactly degenerate) is its steady state.

\subsection{Ultrastrong coupling limit and strong decoherence limit}\label{sec:ultrastrongdyn}

Beyond master equations based on weak coupling expansions,  it is possible to develop \cite{TrushUltra} a dynamical perturbation theory that extends to the ultrastrong coupling regime, see Sec.~\ref{subsec:staticultrastrong}. Consider again the total Hamiltonian \eqref{eq:totalHamiltonian}.
For a non-degenerate $X$ with spectral decomposition into $P_n=\ket{x_n}\bra{x_n}$  given in \eqref{eq:Xspectr}, the trick is to decompose $\HS$ in the pointer basis $\{\ket{x_n}\}$ from the outset, 
\begin{equation}\label{eq:PointerBasisSeparation}
    \HS =\sum_m\varepsilon_m\ket{x_m}\bra{x_m}+\sum_{m\neq n}\Delta_{mn}\ket{x_m}\bra{x_n}.
\end{equation}
Here $\varepsilon_m$ and $\Delta_{mn}$ are real and complex numbers, respectively. 
One may now assume that $\Delta_{mn}$ can be treated as small compared to a very large system-bath coupling strength $\lambda$. More precisely, one should compare $\Delta_{mn}$ with another quantity of the dimensionality of energy. E.g. $\lambda^2J(\omega)$ for a characteristic Bohr frequency $\omega$ may  serve as such quantity. 
Both large $\lambda$ and small $\Delta_{mn}$ imply that the decoherence in the `pointer basis' $\{\ket{x_m}\}$ takes place on a  time scale much smaller than the relaxation of the populations, $p_m=\braket{x_m|\rho|x_m}$, see Ref.~\cite{Merkli2018} for a detailed analysis of the spin-boson model. 
Using this observation, a Pauli master equation for the populations in the pointer basis $p_m=p_m(t)$ is obtained \cite{TrushUltra},
\begin{equation} \label{eq:pops}
    \dot p_m(t)=\sum_{n\neq m} ( k_{mn}p_n(t)-k_{nm}p_m(t) ),
\end{equation}
where the rate constants $k_{mn}$ follow detailed balance 
\begin{equation} \label{eq:detailedbalance}
    k_{mn}=e^{-\beta(\varepsilon_m-\varepsilon_n)}k_{nm},
\end{equation}
and are proportional to $|\Delta_{mn}|^2$ and decrease exponentially \cite{TrushUltra} with $\lambda^2$. 
This is a generalization of the F\"{o}rster approximation, which is well-known in the theory of excitation energy transfer \cite{Mohseni2014}, see Sec.~\ref{subsec:staticultrastrong}. 
The detailed balance condition implies that steady-state populations $p_m$ are proportional to $e^{-\beta\varepsilon_m}$. 
The off-diagonal elements in the pointer basis, $\braket{x_m|\rho(t)|x_n}$, tend to zero for an arbitrary $t>0$ as $\Delta_{mn}\to0$ (or $\lambda\to\infty$). 
Thus, the dynamical steady state at $\lambda \to \infty$ is exactly the static MFG  state $\tmf$ given in \eqref{eq:MFGultrastrong}. 
The proof presented in Ref.~\cite{TrushUltra} extends to   multiple interactions $X^{(j)}$ in \eqref{vint}, degenerate spectra of the $X^{(j)}$, as well as combinations of weak- and strong-coupling parts of the interaction Hamiltonian. All these cases assume strong decoherence between certain subspaces, thus giving the name ``strong decoherence limit'' for the corresponding perturbation theory. The steady state at high, but not infinitely large, $\lambda$ has now also been derived \cite{TrushUltra}, and is found to agree well with the steady state that is numerically solved in Fig.~\ref{fig:equilibration} at strong coupling, see also Subsection \ref{sec:numerical}. Also, numerical calculations of Ref.~\cite{Latune2021} show excellent correspondence between the MFG state and the steady state at high, but not infinitely large $\lambda$.

Pioneering work \cite{Goyal2019, Orman2020}  on the ultrastrong coupling limit has previously conjectured a dynamical steady state based on arguments from einselection theory \cite{Zurek2003}. They argued that the state should be diagonal in the pointer basis $P_m$, which is now confirmed analytically \cite{TrushUltra}. 
The steady state diagonals in the pointer basis were conjectured to be $\langle x_m|\tau |x_m\rangle$ provided that the initial state is the Gibbs state $\tau$.
This conjecture differs from the expressions derived above. Moreover, such state is clearly non-stationary in the case of non-zero $\Delta_{mn}$.

Nevertheless, since $\Delta_{mn}$ are much smaller then $\lambda^2J(\omega)$, decoherence in the pointer basis occurs much faster than the transfer between the different pointer states \cite{TrushUltra}. Hence, the conjectured state is in fact rapidly achieved due to strong decoherence. But thereafter the system still slowly relaxes to the MFG state $\tmf$.  This resolves the discussion~\cite{Cresser2021a,Goyal2019} about the correct form of the steady state at ultrastrong coupling.

Finally, comparing to the mathematical rigorous convergence proof in Sec.~\ref{sect:RtE} for small enough coupling, the proof outlined here for the ultrastrong coupling is only `physically rigorous', i.e. based on a microscopic model and physically plausible assumptions. 
For the spin-boson model, and under certain conditions on the spectral density, the convergence has also been proven mathematically rigorously using a combination of the polaron transformation and quantum  resonance theory,\cite{Konenberg2015,MerkliJMC2016, Konenberg2016} see Subsection~\ref{sec:polarondyn}.

\subsection{Intermediate coupling}\label{sec:StrongCoupling}

So far we have discussed two limiting cases: weak and ultrastrong coupling limits, see Fig.~\ref{fig:weakvsstrong}b). 
But the case of intermediate coupling is highly  interesting. Since in this case,  generally there is no basis for a perturbation theory, non-perturbative techniques need to be developed and employed. We will briefly discuss two known analytical methods that can explore this regime: the reaction coordinate approach and the polaron transformation. 





%
\subsubsection{Reaction coordinate approach}

The reaction coordinate approach is a non-perturbative approach to dealing with intermediate and strongly coupled systems. It was originally developed quite some time ago \cite{Burkey1984,Garg1985} but has recently been much extended to various applications in the quantum thermodynamic context \cite{Hughes2009,Iles-Smith2014,Iles-Smith2016,Strasberg2016,Newman2017,Schaller2018,Nazir2018,Strasberg2018b,Restrepo2018,Newman2020,McConnell2019,McConnell2021,Anto-Sztrikacs2021,Anto-Sztrikacs2021a,Maguire2019,Ivander2021}. 
The basic idea is to redefine that part of the bath $\b$ that contributes most strongly to the system-bath coupling into a collective single degree of freedom, the reaction coordinate (RC). This coordinate is then incorporated along with the original system into an enlarged system $\s \& {\rm RC}$. The remnant degrees of freedom of the original bath $\b$ now constitute a reduced bath $\tilde{\b}$. The whole point is that the enlarged system $\s \& {\rm RC}$ may interact 
only weakly with the reduced bath $\tilde{\b}$, with a new small coupling parameter $\lamBt$. Thus the impact of $\tilde{\b}$ on $\s \& {\rm RC}$ can be treated by the perturbative approaches discussed in the weak coupling section \ref{subsec:weakdyn}. A development of the reaction coordinate method leads to a mapping of a free bosonic bath into a chain of interacting harmonic oscillators coupled to a system at one end \cite{Chin2010a,Prior2010}. This mapping was mentioned in Sec.~\ref{sect:ETH} in relation to many-body physics. It can be used for numerical simulations: the so called time evolving density matrix using orthogonal polynomials (TEDOPA) algorithm.

The reaction coordinate approach can offer insight into the steady state of the original system when the coupling to the original bath is in the intermediate coupling limit, see Fig.~\ref{fig:weakvsstrong}.
First, one obtains an expression for the steady state of the enlarged system $\s \&{\rm RC}$, which is usually just the Gibbs state $\tau_{\s\&{\rm RC}}$ for the corresponding Hamiltonian of $\s\&{\rm RC}$.
By tracing out the reaction coordinate, the steady state of the original system $\s$ is then be found. 
In \cite{Iles-Smith2014}, numerical results are presented that strongly indicate that this state is not the Gibbs state $\tau$. 
In Refs. \cite{Strasberg2016,Nazir2018} the steady state of  $\s$ is shown to be $\tmf+\mathcal{O}(\lamBt)$ where  $\tmf$ is the mean force Gibbs state for $\s$, as defined in \eqref{eq:MFGibbs}. In Ref. \cite{Latune2021a}, steady states obtained by the reaction coordinate method are compared with the perturbative result \eqref{eq:NormalizedMFGSmain} for the MFG state in the weak coupling regime.

\subsubsection{Polaron transformation}
\label{sec:polarondyn}

In Sec.~\ref{sec:polaronstatic} we mentioned the use of the polaron transformation for the derivation of approximations to the mean force Gibbs state. A quantum master equation in the polaron-transformed frame was derived in Refs.~\cite{Brandes2005,Jang2008,Nazir2009,McCutcheon2010,Kolli2011}. Since the polaron transformation mixes the system and bath DoFs, originally, the method allowed to evaluate only the populations (in the system eigenbasis). Formulas for the coherences were derived in Ref.~\cite{Kolli2012}. As in the static considerations, if the off-diagonal elements of the system Hamiltonian in the pointer basis are small enough with respect to the system-bath coupling, the polaron-transformed Bloch-Redfield master equation 
correctly describes the dynamics. 
Again, the variational (instead of full) polaron transformation can be applied; the corresponding master equation and its application to excitation energy transfer and excitonic Rabi rotations in a driven quantum dot were considered in Refs.~\cite{McCutcheon2011,McCutcheon2011a,Pollock2013}

The steady states of the (full) polaron-transformed master equation were studied in Ref.~\cite{Xu2016njp}, see also the  review article~\cite{Xu2016}. They confirm convergence to the mean force Gibbs state.

Combining the polaron transformation and quantum resonance theory, it is shown rigorously in Ref. \cite{Konenberg2015} that the spin-boson model discussed in Subsection~\ref{subsec:staticpolaron} exhibits RtE (see Section~\ref{sec:dynamicsA}) for arbitrary values of the coupling strength $\lambda$. The authors show that under certain conditions on the spectral density (in particular, $J(\omega)$ should be proportional to $\omega^s$ for $s\geq3$ for small $\omega$), if the coupling $\Delta$ in the system Hamiltonian \eqref{eq:spinHS} is smaller than a certain value $\Delta_0$ (depending on the other parameters of the model), then the coupled $\s\b$ dynamics converges to the joint $\s\b$ equilibrium state $\tauglobal$ asymptotically in time. In particular, the reduced state of the system converges to the mean force Gibbs state, which rigorously proves the result of Section~\ref{sec:ultrastrongdyn} for this particular model. In the subsequent works \cite{MerkliJMC2016, Konenberg2016} the reduced dynamics of the population and coherences of the two-level system is  derived for all times.

The analysis of Ref. \cite{MerkliJMC2016} is carried out in a more general setting, including for lower temperatures, for collective  and local (independent donor and acceptor) baths, and for donors and acceptors coupled with different strengths to the bath(s). It leads to a generalized Marcus formula \cite{Marcus1956} with the original one (high temperature, common bath, same donor and acceptor coupling strength) as a special case.

We stress however, that even in the case of proper (super-Ohmic) spectral densities, the polaron transformation followed by weak coupling perturbation theory in the polaron frame cannot serve as a universal tool. It has a certain range of validity and establishing precise conditions on where the weak coupling perturbation theory in the polaron frame can truly be applied, remains largely an open problem.

\subsubsection{Low density limit}

A common assumption in the above discussion on weak coupling dynamics is the choice of interaction \eqref{m18}, which is linear in creation and annihilation operators of the bath.
However, open quantum systems theory also studies other types of interactions, bath models, and limits, such as the singular coupling limit and the low density limit \cite{BrPet2002,Rivas2011}.
While the singular coupling limit is shown  to be equivalent to a special kind of the weak coupling limit \cite{Palmer1977,Accardi1992}, which is captured in the unified weak coupling framework\cite{TrushUni}, the low density limits \cite{Dumcke1985,Accardi2002,Accardi2003} represent significantly different theory. 

Now the interaction $V_{\s\b}$ is quadratic in the creation and annihilation operators, which represents a collision of a system with a particle of the bath. The state of the bath is a grand canonical (instead of canonical) equilibrium state. The low density limit is the limit $n\to0$, where $n$ is the density of particles of the bath. So, the collisions of the system with the particles of the bath are not weak, but rare. 
A close relation of this model to the collision model (or repeated interactions model), popular in the theory of open quantum systems and quantum thermodynamics \cite{Scarani2002,Rybar2012,Barra2015,Bruneau2014a,Zagrebnov2016,Tamura2016,Filippov2017,Ciccarello2017,Cattaneo2021}, is established in Ref.~\cite{Filippov2020}.

In the low density limit, one can derive a Markovian quantum master equation of the GKSL form for the reduced system state $\rho(t)$ \cite{Dumcke1985,Accardi2002,Accardi2003}. The steady state(s) of this equation are much less studied than those for the weak coupling MEs discussed above. Some sufficient conditions for the Gibbs state $\tau$ to be a unique steady state were derived in Refs.~\cite{Dumcke1985,Accardi2019}. However, in a recent paper Ref.~\cite{Accardi2020}, it is shown that, under certain (non-generic) conditions on the spectrum of $\HS$, the system state  converges to a unique steady state that is different from $\tau$. It is an open question whether this steady state is the mean force Gibbs state.

\subsection{Non-perturbative numerical methods} \label{sec:numerical}

There is a range of powerful numerical methods able to solve the dynamics of an open quantum system in the  intermediate coupling regime.  These include the hierarchical equations of motion (HEOM) \cite{Tanimura1989,Tanimura2020} and TEMPO, which is based on time-evolving matrix product operators \cite{Strathearn2018}. 
We will not describe numerical methods here.  
But we used HEOM (in its high-temperature version \cite{Ishizaki2009}) to calculate the dynamics of the qubit system shown in Fig.~\ref{fig:equilibration}. HEOM gives numerically exact results. 
In both the weak and strong coupling cases, we find excellent agreement  of the steady state $\stst$ with the high temperature MFG state formula \eqref{eq:highTMF} from Ref.~\cite{Gelzinis2020}. This is because for the parameters given \cite{figurefootnote} for Fig.~\ref{fig:equilibration}, the temperature is  sufficiently high that even in the strong coupling case the condition $\ell \beta<2$ is still satisfied. 
For the (moderately) strong coupling case in Fig.~\ref{fig:equilibration}, we find very good agreement  of the steady state $\stst$ also with the MFG state formula \eqref{eq:MFGultrastrong} for the  ultrastrong coupling limit ($\lambda \to \infty$) derived in Ref.~\cite{Cresser2021a} Inclusion of corrections to this formula  for  large, but finite, $\lambda$, obtained in Refs.~\cite{TrushUltra,Latune2021}, gives the most precise match with the numerical steady state.




\bigskip

\section{Conclusions and Open questions} \label{sec:conclusions}

We now conclude on the findings outlined above and provide an extensive list of open questions. 

Sec.~\ref{sec:statics} summarised the static point-of-view. This view includes non-negligible coupling into a modified equilibrium state, the MFG state $\tmf$, and provides the backbone for much current research on constructing a   thermodynamic framework  that goes beyond Gibbs state physics \cite{Jarzynski2004a,Campisi2009a,Campisi2009b,Gelin2009,Hilt2011,Hilt2011a,Seifert2016,Philbin2016,Jarzynski2017,Aurell2017,Strasberg2017b,Miller2017,Aurell2018,Miller2018a,Schaller2018,Strasberg2018b,Miller2018,Strasberg2019,Correa2017,Hovhannisyan2018a,Perarnau-Llobet2018}.
Despite the growing importance of this framework for  the assessment of thermodynamic processes of nanoscale and quantum systems, e.g., in terms of heat and work exchanges as well as entropy production on the level of fluctuations and averages, explicit expressions for the concrete functional shape of $\tmf$, are only known in a handful of cases. 

Beyond the exactly solvable quantum oscillator, explicit expressions for general systems are known for weak (neglecting ${\cal O}(\lambda^4)$ and higher) and ultrastrong  (neglecting ${\cal O}(\lambda^{-1})$  and higher)  coupling limits. 
It remains an open question \Qu{\QuNo} as to whether or not there exists relatively simple, tractable expressions for the MFG state for intermediate coupling regimes, at least for specific system choices. This is the interesting regime where, loosely speaking, the system’s interaction with its environment is of comparable scale to the system’s bare energy. For example, it may be possible to construct useful expansions for $\lambda$ that neither expand for small or large $\lambda$, but some intermediate value $\lambda_0$; but we note that previous considerable analytical effort has not resulted in `simple' expressions even for classical systems.
\footnote{For many numerical methods that can explore this regime, finding the long time steady state can be improved by repeated shorter time evolution \cite{Purkayastha2021}. 
Furthermore, a recent paper \cite{Chiu2021} discusses an efficient numerical method based on tensor-network states in imaginary time that can compute the $\tmf$ at arbitrary coupling strength $\lambda$ and temperature.}
It also remains open \Qu{\QuNo} whether there exists a general criterion, or criteria, that allows one to judge, just looking at the Hamiltonian  of a particular problem, whether mean force corrections are going to be important or not in the equilibrium state.  
For the weak coupling limit such  criteria exist (see e.g. \eqref{eq:binomialApproxConditionmain}), which reveal a complex interplay of system energies, system-bath coupling parameters, and bath temperature. Analogous condition(s) in other coupling regimes remain to be uncovered.

Another open question is the extension of the above results, all valid for coupling to a single (continuum) bosonic bath, to \Qu{\QuNo} simultaneous coupling to multiple bosonic baths and to \Qu{\QuNo} fermionic baths. \Qu{\QuNo} Non-linear couplings in the bath ladder operators could also be considered. 

Furthermore, connections between quantum and classical MFG states are well worth exploring. Firstly they would \Qu{\QuNo} provide a direct connection \cite{Jarzynski2017} to numerical simulations in (classical) chemistry \cite{Roux1995,Maksimiak2003,Allen2006,Lahey2020}. These routinely include potential of mean force corrections in the calculation of arrangements of molecules in solutions in a vast range of contexts. A timely example are simulations of the catalytic mechanism of proteases in respiratory syndromes \cite{Wang2020}. 
Secondly,  \Qu{\QuNo}  a comparison between quantum and classical cases could also illuminate  quantum signatures in the mean force Gibbs state that are not present in the classical counter part \cite{Berritta2021}.
Another  intriguing question \Qu{\QuNo} is whether one can derive, or not, the MFG state $\tmf$ in some manner from entropy-maximisation arguments in line with such derivations of the standard Gibbs state, as outlined at the start of section \ref{sec:statics}. For both the classical or quantum case, the difficulty is here that the strong-coupling corrected system energy and entropy functionals can depend on temperature \cite{Seifert2016,Jarzynski2017,Miller2018a,Strasberg2019a}.

\smallskip

Sec.~\ref{sec:dynamicsA} outlined how equilibrium arises dynamically, culminating in the mathematical proof that a $\s\b$ complex does equilibrate exactly to the {\it global} Gibbs state $\tauglobal$, when the initial state is not too far away from it,  and the coupling $\lambda$ is finite and small. Under those assumptions, this result immediately dynamically justifies the postulated $\tmf$ used in the statics section.  
A major open question is to \Qu{\QuNo} give a detailed quantification of the upper bound on  $\lambda$ in terms of physical quantities, and \Qu{\QuNo} to prove  the global convergence to $\tauglobal$ and/or the local convergence \footnote{We are aware of work in preparation by G. Guarnieri, A. Purkayastha, J. Anders, {\it et al.}, on a (physically rigorous) proof of dynamical convergence of the system state to $\tmf$ at arbitrary coupling strength $\lambda$.} 
to $\tmf$ for a larger (or the whole) range of coupling strengths $\lambda$. 
Related to this, all bounds obtained so far shrink with temperature $T$. It remains an open problem to \Qu{\QuNo} show RtE at  low (or zero) temperatures
\footnote{For zero temperature, probably, the pseudomode method\cite{Garraway1997,Garraway1997a,Pleasance2020,Pleasance2021,Tere2019,Tere2019a}, which reduces the exact system dynamics to a Markovian dissipative dynamics of a system added by a finite number of harmonic modes, might be helpful.} $T \ll \lambda^2$.  

Another issue is dimension: in all proofs of RtE, the system $\s$ is assumed to be finite-dimensional. Giving \Qu{\QuNo} a proof of RtE for infinite-dimensional systems (with discrete spectrum), such as a harmonic oscillator, is an important  problem to solve. Likewise for an environment with a discrete spectrum,
\Qu{\QuNo} estimating the effective dimension $d_{\rm eff}$ in Eq.~\eqref{eq:deff}  in terms of the volume and/or the number of bath modes
is an open problem.

Finally,  current studies  in quantum thermodynamics  often make use of the properties of  CPTP maps and define the system's dissipated heat $Q$ (with implications for work, entropy, efficiency etc.) as the energy received by an environment \cite{Esposito2010} during a global unitary operation on an $\s\b$ complex. But if the environment is small, such as a qubit, then the very notion of it being a thermodynamic context is being abandoned. If the environment is somewhat larger, such as a discrete `bath'  of harmonic oscillators, then this is somewhat reasonable given the results discussed in \ref{subsect:effdim}. But it is worth keeping in mind that even in this case the dynamics doesn't produce the irreversible character, see Sec.~\ref{sect:RtE}, often assumed in thermodynamic arguments.  A major task is \Qu{\QuNo} the careful assessment of how the results outlined in Sec.~\ref{sec:dynamicsA} come to bear on such thermodynamic characterisations.
 
\smallskip

Sec.~\ref{sec:dynamicsB} outlined a selection of microscopic approaches, in particular master equations, which are capable of giving analytical details of a system's evolution as well as an indication of its steady state. While the application of the secular approximation in the weak coupling MEs of Davies and others predict $\tau$ as steady state, the Bloch-Redfield ME predicts  $\lambda^2$-corrections to $\tau$. However here the issue was that not all  $\lambda^2$-corrections are in fact captured by the  $\lambda^2$-order BRME: some corrections require higher order expansions. 

Most ME treatments assume linear coupling to bath modes.  However, in the low density limit, quadratic (collision-like) couplings \cite{Accardi2020} are important and  give rise \Qu{\QuNo} to open problems concerning their steady states and  consistency with $\tmf$.
In the intermediate coupling limit, reaction coordinate or polaron transformation methods can be used, but they cannot serve  as universal tools as they have their own limitations.
Finally,  the ultrastrong coupling regime stands out because here the ME discussed in Sec.~\ref{sec:ultrastrongdyn}  leads, without any ambiguities, to the corresponding mean force Gibbs state \eqref{eq:MFGultrastrong} as detailed in Sec.~\ref{subsec:staticultrastrong} (which differs from the standard Gibbs state $\tau$).

\smallskip

To conclude, in this article we provided an introduction to several current avenues in the disparate fields of open quantum systems, strong coupling thermodynamics and beyond. 
The aim of these fields is to uncover the bath's signature on a nanoscale system's equilibrium state, as well as elucidate the system's approach to equilibrium.
Impressive results have been achieved addressing this timely challenge -- but many key questions remain open. 
Solving these will provide much needed clarity on how to consistently characterise the thermodynamics of nanoscale and quantum systems. 

While some researchers may feel it is obvious that the MFG state $\tmf$ should be the equilibrium state, we highlight that much current research in quantum thermodynamics still tacitly assumes it to be the Gibbs state $\tau$. This includes many master equation derivations as well as the theory of thermal operations and thermodynamic resource theory to name a few. 
Causes for this scientific mismatch might be 
the unwieldy formal definition of the $\tmf$, partially resolved in Sec.~\ref{sec:statics}, and the conflict between unitary evolution and irreversibility, partially resolved in Sec.~\ref{sec:dynamicsA}. 
The results outlined above provide a glimpse of a theory that goes beyond Gibbs statistical physics, and will find applications in a variety of fields where the exchange of energy on the nanoscale is essential, from quantum chemistry and biology, to magnetism and nanoscale heat management.

\begin{acknowledgments}

We  thank Ahsan Nazir, Chris Jarzynski, Dvira Segal, and Jonathan Keeling for valuable comments on a draft of the manuscript. We are indebted to Philipp Strasberg for  extensive and thoughtful comments, and for bringing up the Gaussian limit argument that underpins much of the success of the Caldeira-Leggett model. 
MM thanks Gennady Berman, J\"urg Fr\"ohlich, Alain Joye, Martin K\"onenberg and Michael Sigal, for guidance and collaborations over many years. AT is grateful to Alexander Pechen, Alexander Teretenkov, Camille Lombard Latune and Oleg Lychkovskiy for valuable comments, bibliography suggestions and fruitful ongoing discussions. 
JDC and JA thank Simon Horsley, Marco Berritta, Stefano Scali and Federico Cerisola for many stimulating discussions on the subject of this  article.
MM is supported by a {\em Discovery Grant} from the {\em Natural Sciences and Engineering Research Council of Canada} (NSERC). 
JA and JDC acknowledge funding from the Engineering and Physical Sciences Research Council (EPSRC) (EP/R045577/1). 
JA further acknowledges EPSRC support in form of a Doctoral Training Grant, and thanks the Royal Society for support.

\end{acknowledgments}

\smallskip

{\it Author Declaration:} The authors have no conflicts to disclose.

\smallskip

{\it Data Sharing Statement:}
The dynamical evolution data displayed in Fig.~\ref{fig:equilibration} are available upon reasonable request to~AT.

\smallskip

\bibliographystyle{apsrev4-1.bst}
\bibliography{OpenSysDynMFG.bib}

\end{document}